\documentclass[fleqn,usenatbib]{mnras}
\pdfoutput=1
\usepackage{newtxtext,newtxmath}
\usepackage[T1]{fontenc}

\DeclareRobustCommand{\VAN}[3]{#2}
\let\VANthebibliography\thebibliography
\def\thebibliography{\DeclareRobustCommand{\VAN}[3]{##3}\VANthebibliography}

\usepackage{graphicx}	% Including figure files
\usepackage{amsmath}	% Advanced maths commands
\usepackage{amssymb}	% Extra maths symbols
\usepackage{threeparttable}
\usepackage{xcolor}

\title[Search for Masers in Dwarf Galaxies]{A search for H$_2$O masers in 100 active dwarf galaxies}

\author[Rosenthal \& Zaw]{
M. J. Rosenthal$^{1}$\thanks{E-mail: mjr612@nyu.edu} and
I. Zaw$^{1}$
\\
$^{1}$Department of Physics, New York University Abu Dhabi, P.O. Box 129188, Abu Dhabi, United Arab Emirates}

\date{Accepted 2020 August 28. Received 2020 August 27; in original form 2020 August 09}

\pubyear{2020}

\begin{document}
\label{firstpage}
\pagerange{\pageref{firstpage}--\pageref{lastpage}}
\maketitle

\begin{abstract}

We present the results of the first dedicated survey for 22 GHz H$_2$O maser emission in dwarf galaxies outside of the local group, with the aim of discovering disc masers. Studies of disc masers yield accurate and precise measurements of black hole mass, and such measurements in dwarf galaxies would be key to understanding the low-mass end of BH--galaxy co-evolution. We used the Green Bank Telescope to survey 100 nearby ($z\lesssim0.055$) dwarf galaxies ($M_*\lesssim 10^{9.5}~M_{\sun}$) with optical emission line ratios indicative of accretion onto a massive black hole. We detected no new masers down to a limit of $\sim$12 mJy (5$\sigma$). We compared the properties of our sample with those of $\sim$1,850 known detections and non-detections in massive galaxies. We find, in agreement with previous studies, that masers are preferentially hosted by Seyferts and highly-obscured, [O {\footnotesize III}]-bright AGNs. Our sample has fewer Seyferts, is less obscured, and is [O {\footnotesize III}]-faint. Though the overall maser detection rate is $\sim$3\% in massive galaxies, the predicted rate for our sample, weighted by its optical properties, is $\sim$0.6--1.7\%, corresponding to a probability of making no detections of $\sim$20--50\%. We also found a slight increase in the detection rate with increased stellar mass in previously surveyed galaxies. However, further observations are required to discern whether there is an intrinsic difference between the maser fraction in active dwarf galaxies and in their massive counterparts for the same AGN properties.

\end{abstract}


%%%%%%%%%%%%%%%%%%%%%%%%%%%%%%%%%%%%%%%%%%%%%%%%%%

%%%%%%%%%%%%%%%%% BODY OF PAPER %%%%%%%%%%%%%%%%%%

\section{Introduction} \label{sec:intro}

There is now an abundance of observational evidence that supermassive black holes (SMBHs) with masses of $M_{\rm{BH}}\sim10^6$--$10^{10} M_{\sun}$ occupy the centres of all massive galaxies with total stellar masses of $M_* \gtrsim 10^{10.5} ~ M_{\sun}$ \citep[e.g.,][]{KormendyHo13}. The presence of SMBHs with $M_{\rm{BH}}\sim10^9~M_{\sun}$ at high redshift \citep[e.g., ][]{Banados18} indicates that massive black holes (BHs) must have grown rapidly in the early universe; yet there is currently little observational evidence to constrain formation scenarios for the first ``seeds'' of today's massive BHs \citep[e.g.,][]{Volonteri10, Greene12}. Empirical scaling relations between BH mass and host galaxy properties such as stellar velocity dispersion ($\sigma_*$), bulge stellar mass ($M_{\rm{Bulge}}$), and $M_*$ suggest a co-evolution of SMBHs and their hosts, but these are largely based on measurements of nearby, massive, and mostly quiescent galaxies \citep[e.g.,][]{FerrareseMerritt2000, Gultekin09, KormendyHo13, McconnellMa13, ReinesVolonteri15}.

Recent numerical simulations of galaxy formation predict limited BH growth for low-mass galaxies, until their hosts reach bulge masses of $M_{\rm{Bulge}}\sim10^{10}~M_{\sun}$ or total stellar masses of $M_*\sim10^{10.5}~M_{\sun}$, resulting in breaks on the associated scaling relations \citep[e.g.][]{AnglesAlcazar17, Bower17, Dekel19}. If massive BHs hosted in galaxies lighter than these critical masses lie either at or below the masses predicted by the scaling relations, then dwarf galaxies with $M_*\lesssim10^{9.5}~M_{\sun}$ should host intermediate-mass black holes (IMBHs) with masses $M_{\rm{BH}}\lesssim10^5~M_{\sun}$. Observations of IMBHs in low-mass galaxies are required to test the theoretical predictions about the low-mass end of BH and galaxy coevolution.

Despite the ubiquity of their massive counterparts, however, there remain few observational constraints on IMBHs \citep[e.g.,][]{Greene19}. Most measurements of $M_{\rm{BH}}\lesssim5\times10^5~M_{\sun}$ to date have been Virial measurements from single-epoch spectroscopy of broad H$\alpha$ or H$\beta$ emission, which have systematic errors as high as 0.5 dex \citep[e.g.,][]{Reines13, ReinesVolonteri15, Chilingarian18, Baldassare20}. More precise measurements have been accomplished with reverberation mapping for a handful of objects, such as the $\sim$10,000 $M_{\sun}$ BH in NGC 4395 \citep{Woo19}. Both methods, however, require the presence of an unobscured broad line region (BLR), restricting measurements to Type 1 active galactic nuclei (AGNs), and are calibrated to the fiducial $M_{\rm{BH}}$--$\sigma_*$ relation. A few IMBHs have also been detected in quiescent galaxies through stellar or gas dynamical modelling \citep[e.g.,][]{Nguyen17}, but spatial resolution requirements restrict such measurements to a small sample of very nearby galaxies. BH masses in Type 2 AGNs with obscured BLRs remain mostly un-probed in the low-mass regime.

Water megamaser emission at 22 GHz is the only known tracer of both the position and velocity of warm, dense gas \citep{Neufeld94} in the central $\sim$0.1--1 pc of nearby ($z\lesssim0.06$) AGNs. In some systems, maser emission arises from a circumnuclear disc with dynamics dominated by the gravity of the central BH. Fitting Keplerian curves to the positions and velocities of the emission in such systems yields the most accurate measurements of BH mass outside the local group \citep[e.g.,][]{Miyoshi95, Kuo11, Humphreys13}. Due to the requirement of long amplification gain paths, maser discs must be viewed nearly edge-on, and thus are primarily detected in Type 2 AGNs \citep[e.g.,][]{Zhang06, Greenhill08, Zhang10, Castangia19}, making maser BH mass measurements ideal complements to BLR-based techniques. The use of water masers to measure BH masses is limited by the rarity of viable systems -- water maser emission is found in only $\sim$3\% of galaxies out to $z\lesssim0.05$, and only $\sim$1/3 of these are disc masers, with another $\sim$1/3 associated with jets, and the remainder of unknown physical origin \citep{Zhu11, Constantin12, Kuo18}.

Until recently, all maser-measured BH masses were of $M_{\rm{BH}}\geq10^6~M_{\sun}$. However, new observations of maser emission in the low-mass galaxy IC 750 ($M_*=10^{10.1\pm0.2}~M_{\sun}$) with the Very Long Baseline Array (VLBA) find evidence for an IMBH in the nucleus, and measure a dynamical mass upper limit of $M_{\rm{BH}}\leq1.4\times10^5~M_{\sun}$ from emission within a $\sim$0.2 pc diameter disc \citep{Zaw20}. This mass falls 1--2 dex below that predicted by existing scaling relations, but it is unclear as of yet whether this is due to larger scatter at the low-mass end of the scaling relations, or because masers in low-mass galaxies trace BH masses systematically below the lines fit to massive galaxies, as seen in some simulations \citep[e.g.,][]{AnglesAlcazar17, Bower17, Dekel19}. A systematic search for additional disc masers in low-mass galaxies is necessary to distinguish between these scenarios.

To date, the majority of surveys for extragalactic water masers have been in massive galaxies \citep[e.g.,][]{Braatz04, Kondratko06a, Castangia08, vandenBosch16}. The only dedicated survey for 22 GHz water masers in dwarf galaxies targeted 17 galaxies in the local group, and resulted in no new detections \citep{Tarchi20}. The largest available sample of active dwarf galaxies is from optical spectroscopic measurements by \citet{Reines13}, who identify 136 dwarf galaxies with optical emission line ratios suggesting accretion onto a massive BH. The majority of these galaxies have not previously been surveyed for maser emission.

In this paper we report the first dedicated search for water maser emission in active dwarf galaxies beyond the local group, with the Robert C. Byrd Green Bank Telescope (GBT), targeting galaxies from the \citet{Reines13} sample. In Section \ref{sec:sample} we describe the selection of our target list from this parent sample. We describe our observations with the GBT, and our data reduction methods in Section \ref{sec:data}. In Section \ref{sec:results} we present the results of our survey, and give the upper limit on the maser detection fraction in dwarf galaxies based on our lack of detections. In Sections \ref{sec:cross-match}--\ref{sec:properties}, we examine the detection rate in previous surveys as a function of target galaxies' optical properties. We conclude in Section \ref{sec:conclusion}. All calculations in this paper assume a Hubble constant of $H_0 = 70$ km s$^{-1}$ Mpc$^{-1}$.

\section{Sample Selection} \label{sec:sample}

We select our targets from a parent sample of 136 active dwarf galaxies from \citet{Reines13}. These were identified from $\sim$26,000 galaxies with $\log(M_*/M_{\sun})\lesssim9.5$ from the NASA-Sloan Atlas\footnote{v0.1.2; \url{http://www.nsatlas.org/}} (NSA), a redshift-limited ($z\leq0.055$) sample of galaxies based on the spectroscopic catalog from SDSS DR8 \citep{York2000, Aihara11}. \citet{Reines13} report 35 AGNs and 101 composites, based on their optical emission line ratios.

 \citet{Reines13} report 35 AGNs with [O {\footnotesize III}] $\lambda$5007 / H$\beta$ and [N {\footnotesize II}] $\lambda$6583 / H$\alpha$ emission line ratios above the theoretical upper limit for photoionization by star-formation from \citet{Kewley01}, and an additional 101 composites with emission line ratios below this limit, but above the empirical AGN diagnostic line from \citet{Kauffmann03}.

To restrict our targets to Type 2 nuclei, we exclude 6 AGNs and 4 composites in which \citet{Reines13} detect broad H$\alpha$ emission. We also exclude 18 AGNs and 9 composites that had already been surveyed for 22 GHz maser emission with the GBT, as shown on the list of targets maintained by the Megamaser Cosmology Project\footnote{\url{https://safe.nrao.edu/wiki/bin/view/Main/MegamaserCosmologyProject}} (MCP). We are left with 99 galaxies, of which there are 11 AGNs and 88 composites.

We also surveyed Was 49b, a member of a dual-AGN system in an active merger \citep{Secrest17}. Spectropolarimetry of Was 49b shows broad lines in its polarized optical spectrum, similar to those of known disc masers such as NGC 1068 \citep{Greenhill96} and Circinus \citep{Greenhill03}. This brings the total sample size to 100 galaxies that we target in our observations, and 27 additional galaxies with archival GBT spectra.

\section{Observations and Data Reduction} \label{sec:data}

Observations were taken using the K-band Focal Plane Array (KFPA) on the GBT between November 7, 2019, and February 9, 2020, under project code AGBT19B\_281. We used Mode 5 of the Versatile GBT Astronomical Spectrometer (VEGAS), which has an instantaneous bandwidth of 187.5 MHz and spectral channel width of 2.9 kHz, or $\sim$0.08 km s$^{-1}$ at 22 GHz. As the largest known velocity offset for water maser emission is $\sim$1000 km s$^{-1}$ from the systemic \citep[e.g. NGC 4258,][]{Miyoshi95}, we observed with four spectral windows, with one centred at the frequency of the water maser line in frame of the heliocentric systemic velocity of the galaxy, and the other three offset by --150 MHz, 150 MHz, and 300 MHz, sufficient to cover the most offset maser emission.\footnote{We observed with only one spectral window, centred on the systemic velocity of the target, nodding with beams 4 and 6 for our first observation on November 7, 2019 (session AGBT19B\_281\_01), due to a configuration issue. This issue was corrected, and all subsequent observations used four spectral windows and nodded between beams 2 and 7.} Heliocentric velocities and equatorial coordinates for each target were taken from NED.\footnote{\url{https://ned.ipac.caltech.edu/}} Each target was observed with two consecutive dual-polarization, dual-beam nods for a total integration time of 10 minutes.

We conducted pointing and focus calibration at the beginning of each session and once every $\sim$1 hour thereafter, corresponding to 4 or 5 targets between pointings depending on slew time. Once per session, to check our setup, we also observed one known H$_2$O maser. To ensure clear detections without significant additions to the overhead, we only targeted known masers with strong ($\gtrsim$100 mJy) and broad ($\gtrsim$10 km s$^{-1}$) emission, such that they could be detected by single nods of under 5 minutes of integration. These galaxies were NGC 4258, NGC 3079, NGC 1068, and IC 485.

We averaged the nods and polarizations for each target to increase signal to noise (S/N)\footnote{Maser emission from AGN is unpolarized, so averaging polarizations has no effect apart from increasing S/N.}. When spectra contained radio-frequency interference (RFI), we blanked the affected channels. We then boxcar smoothed the spectra by 16 channels, the lowest power of 2 for which smoothed channel widths were $>$1 km s$^{-1}$, the width of a maser line. We fit the line-free regions of the smoothed spectra with fifth-degree polynomials, which we subtracted from the spectra before finally Hanning smoothing. We measured the root-mean-square (RMS) flux density from emission-free regions both before and after Hanning smoothing. The mean RMS for the sample is 3.69 mJy before Hanning, and 2.31 mJy after Hanning. Smoothed channel widths were 1.2--1.4 km s$^{-1}$, depending on the systemic velocity of the galaxy.

It is typical for GBT spectral line data reduction to involve smoothing the off-source spectrum to boost S/N. Using \texttt{smthoff} 16 in this case caused an apparent increase in the rate of RFI in our observations, so we did not smooth the off spectrum for our primary data reduction. In some cases, narrow emission features were initially detected, but when re-examined with varying \texttt{smthoff} settings in GBTIDL, were revealed to be RFI.

Additionally, we re-reduced the archival GBT data for the 27 Type 2 AGNs in dwarf galaxies from \citet{Reines13} that were previously observed by the MCP. Some of these observations used different VEGAS modes, or used the GBT spectrometer from before VEGAS was implemented, and thus have different unsmoothed channel widths. In all cases we boxcar smoothed by the lowest power of 2 such that smoothed channels were $>$1 km s$^{-1}$ in width, and otherwise used an identical procedure as described above. The mean RMS for the archival data is 5.35 mJy before Hanning, and 3.38 mJy after Hanning.

\section{Survey Results} \label{sec:results}

We detect no maser emission in any of the 100 dwarf galaxies in our survey above 5$\sigma$, which corresponds to a mean detection threshold of $\sim$12 mJy. The lack of a detection in any given galaxy could be for one of three reasons: 1) The galaxy does not host maser emission at all; 2) the galaxy hosts maser emission that is always below our detection threshold; or 3) because masers are variable on time scales of months \citep[e.g.][]{Castangia08, Humphreys13} the galaxy hosts maser emission that is below our detection threshold during these observations, but could be detectable at a different time. In our initial data reduction, there was one apparent detection in the galaxy SDSS J114359.58+244251.7 (RGG 18), but closer inspection revealed that the signal was strongly polarized, and therefore not real maser emission. Details for the non-detections from our survey are given in Appendix \ref{sec:nondetections}.

The overall detection rate for masers in all galaxies is $\sim$3\% \citep[e.g.,][]{Zhu11, Constantin12}, based on previous surveys, which mostly targeted massive active galaxies. Assuming a simple binomial distribution, the probability of having zero detections among 100 galaxies with an assumed rate of $\sim$0.03 is $\sim$5\%. Alternatively, the 95\% confidence level (CL) upper bound on zero counts assuming Poisson statistics is 2.996 \citep{Gehrels86}, or a 3.0\% upper bound on the detection rate, meaning we cannot rule out that the intrinsic rate of H$_2$O masers among dwarfs is consistent with that for more massive galaxies.

There are also no detections in the archival data. Including these non-detections brings the total sample size to 127. The binomial probability of no detections with an assumed rate of 0.03 is 2.1\%, and the 95\% CL upper bound for the intrinsic rate is 2.4\%. In computing bounds on the detection rate while including the archival data, however, we note that the mean RMS noise values for the archival data, 5.35 mJy after boxcar smoothing and 3.38 mJy after boxcar and Hanning smoothing, are $\sim$50\% higher than in our new observations.

While it is plausible that our lack of detections is simply due to small sample size rather than intrinsic differences in the maser fraction among dwarf galaxies and their more massive counterparts, it is worthwhile to compare the properties of the galaxies in our sample against galaxies from previous surveys. It is well-documented that maser detection rates vary with the radio \citep[e.g.][]{Zhang17}, infrared \citep[e.g.][]{Kuo18, Kuo20}, optical \citep[e.g.][]{Zhu11, Constantin12}, and X-ray \citep[e.g.][]{Greenhill08, Zhang10, Castangia19, Kuo20} properties of surveyed galaxies.

All of our targets except for Was 49b, which has $z>0.055$, are from \citet{Reines13}, and all have NSA IDs. We can therefore compare the remaining 99 dwarf galaxies with more massive NSA galaxies that have been surveyed for water masers with the GBT. In the remainder of this paper, we identify a sample of galaxies are in the NSA and in the MCP's list of GBT observations. We then examine the maser detection rate in previously surveyed NSA galaxies as a function of their optical properties, and compare the properties of the 99 newly-surveyed dwarf galaxies with those of their massive counterparts.

\section{Identifying GBT-Surveyed Galaxies with NSA Spectroscopy} \label{sec:cross-match}

\subsection{GBT Observations from the MCP}

% Address the fact that not all confirmed masers are GBT observations, so neither list is a subset of the other

The MCP maintains two catalogs pertaining to the search for extragalactic 22 GHz water masers: a list of 6,355 GBT observations at 22 GHz which includes both detections and non-detections\footnote{\url{https://safe.nrao.edu/wiki/bin/view/Main/MegamaserProjectSurvey}. The 6,355 GBT observations are inclusive of the 27 previously-surveyed active dwarf galaxies from \citet{Reines13} that we did not re-observe.}, and a catalog of 180 confirmed H$_2$O maser host galaxies\footnote{\url{https://safe.nrao.edu/wiki/bin/view/Main/PublicWaterMaserList}.}. From these catalogs we identify a sample of GBT detections and non-detections that have cross-IDs in the NSA.

The majority of galaxies in these catalogs, specifically 96\% of the GBT observations and all but five of the confirmed detections, are within the NSA redshift limit of $z\leq0.055$ ($v_{\rm{sys}} \leq 16489$ km s$^{-1}$). We remove galaxies that do not satisfy this requirement, as well as 21 observations with a recorded velocity of 0 km s$^{-1}$. Of the 175 detections within the redshift limit, 22 are from non-GBT surveys, and do not correspond to any of the 6,355 GBT observations listed by the MCP, and we remove these. A further 7 of the remaining 153 are confirmed or suspected to be associated with star-forming regions rather than AGNs. Removing these, we are left with a sample of 146 unique galaxies with GBT maser detections, which form a subset of the 6,355 total GBT observations.

% Change the number of unique galaxies to include the redshift limit and v = 0 limit. And change the rate to reflect both this and the fact that the number of good detections is lower than 161 (at least 4 removed for redshift, several others for not being GBT).

While each of the 146 detections is a unique galaxy, the larger list of observations includes many galaxies with truncated names, names that are abbreviations of their IAU names, and a number of galaxies with duplicate observations. In some cases, repeated observations of the same galaxy use different naming conventions. We remove all duplicates, leaving 4,677 unique galaxies with $z\leq0.055$ (146 detections and 4,531 non-detections). The unfiltered detection rate is $3.12\pm0.26\%$, using Gaussian errors.

\subsection{Matching the MCP and NSA samples}

Non-standard naming conventions and truncations prohibit us from resolving source names for a significant fraction of the GBT observations listed by the MCP. Instead, to match GBT-surveyed galaxies to the NSA, we perform a simple nearest-neighbor association. We imposed a maximum separation of 3''.3, the angular separation at which $A_e/A_0 = 90\%$ for the GBT beam at 22 GHz. Of the 4,677 GBT-surveyed galaxies with $z \leq 0.055$, 3,206 are matched with an NSA galaxy. Since the 27 dwarf galaxies from \citet{Reines13} with archival GBT spectra are part of both the NSA and MCP samples, they are included in these 3,206 galaxies.

To prevent association of GBT targets with foreground or background NSA galaxies, we impose a limit on the difference in recessional velocity, $|cz - v_{\rm{sys}}|\leq100$ km s$^{-1}$, where $v_{\rm{sys}}$ is the heliocentric recessional velocity from the GBT observation, $c$ is the speed of light, and $z$ is the heliocentric redshift from the NSA. Of the 3,206 matched galaxies, 2,975 (93\%) satisfy this requirement. Additionally, our analysis in Sections \ref{subsec:BPT}--\ref{subsec:lumandmass} will rely on measurements of the [O {\footnotesize III}] $\lambda5007$, [N {\footnotesize II}] $\lambda6583$, H$\alpha$, and H$\beta$ emission lines, so we restrict our sample to galaxies with $\rm{S/N}\geq3$ for these four lines. Of the 2,817 initial identifications, 2,002 galaxies (62\%) satisfy the S/N requirement, and 1,847 galaxies (58\%) satisfy both the velocity and S/N requirements.

Of the 146 GBT-detected known masers with $z \leq 0.055$ that are not suspected to be associated with star formation, 97 are recovered as a subset of the 3,206 GBT target galaxies in the unfiltered, matched NSA and MCP catalogs, for a rate of $3.02\pm0.31$\% (assuming Poisson errors on detection counts). After filtering for velocity matching and good spectroscopy, there remain 46 detections out of 1,847 MCP-surveyed galaxies, for a rate of $2.49\pm0.37$\%, indicating the detection rate is not strongly biased in the subsample with good spectroscopy. A summary of the samples we use and what cuts are applied, as well as the number of galaxies and detections after each cut, is shown in Table \ref{table:sample}. Henceforth we use ``MCP--NSA sample'' to refer to the matched sample of 1,847 galaxies with velocity matching and good spectroscopy.

The majority of the galaxies in the MCP--NSA sample have source names in their MCP documentation that were resolvable in NED. For each of these, we checked that resolving the object's NSA ID and its MCP name returned the same results, and found no false positive matches.

\begin{table*}
    \centering
    \caption{Description of the samples used in this work}  \label{table:sample}
    \begin{tabular}{lcc}
        \hline
        \hline
         & $N_{\rm{galaxies}}$ & $N_{\rm{detections}}$ $^a$ \\
	\hline
	\citet{Reines13} dwarf galaxies & 136 & 0 \\
	... of which are narrow-line AGNs or composites & 126$^b$ & 0 \\
	... of which are narrow-line AGNs or composites and were previously surveyed with the GBT & 27 & 0 \\
	... of which are surveyed for the first time in this work & 99$^c$  & 0 \\
	\hline
	MCP unique GBT-surveyed galaxies with $z\leq0.055$ & 4677$^d$ & 146$^e$ \\
	\hline
        MCP unique GBT-surveyed galaxies, $z\leq0.055$, matched with NSA, $\theta_{\rm{sep}}<3.3$'' & 3206$^d$ & 97$^e$ \\
        ... with good velocity agreement: $|v - cz|<100$ km s$^{-1}$ & 2975 & 89 \\
        ... with good velocity agreement and $\rm{ZDIST/ZDIST\_ERR}>10$ & 2877 & 78 \\
        ... with good spectroscopy: $\rm{S/N(H\alpha,H\beta,[N~II],[O~III])}>3$ & 2002 & 53 \\
        ... with both good velocity agreement and good spectroscopy & 1847 & 46 \\
        ... with good velocity agreement, good spectroscopy, and $\rm{S/N([S~II])} > 3$ & 1835 & 46 \\
        ... with good velocity agreement, good spectroscopy, and $\rm{ZDIST/ZDIST\_ERR}>10$ & 1818 & 42 \\
        \hline
    \end{tabular}
    \tablenotes{
    \item $^{a}$ Number of galaxies with detected circumnuclear maser emission.
    \item $^{b}$ Classified as narrow-line AGN or composite based on the lack of a broad H$\alpha$ detection and the emission line ratios measured by \citet{Reines13}. 
    \item $^{c}$ Our target sample consists of these 99 dwarf galaxies, and Was 49b, which lies just outside the redshift limit of the NSA. 
    \item $^{d}$ Includes the 27 galaxies with narrow-line AGNs from \citet{Reines13} that were previously surveyed with the GBT.
    \item $^{e}$ Does not include galaxies where maser emission is known or suspected to be associated with star-formation rather than an AGN.} 
\end{table*}

\begin{figure*}
    \centering
    \includegraphics[width=\linewidth]{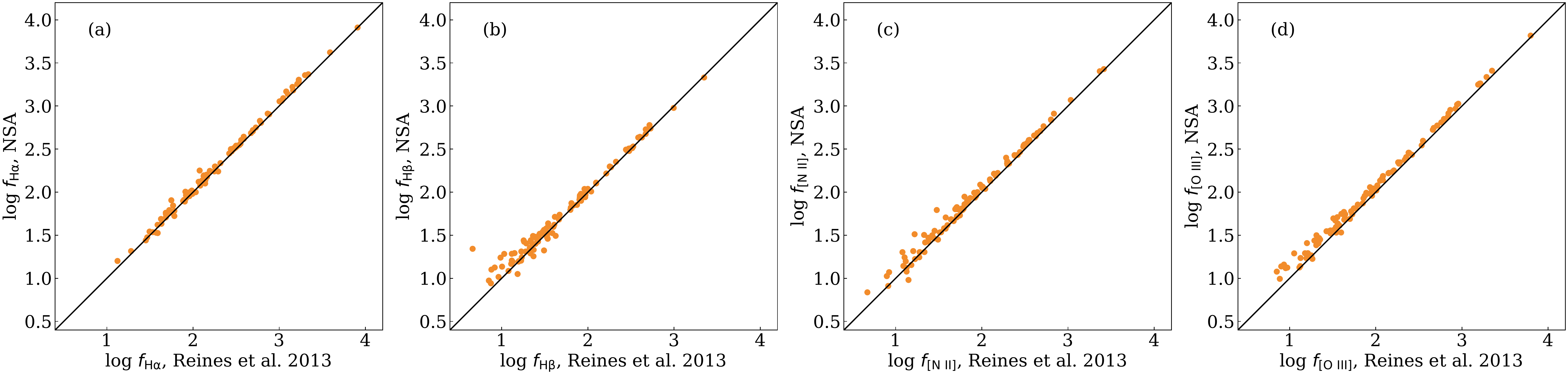}
    \includegraphics[width=0.75\linewidth]{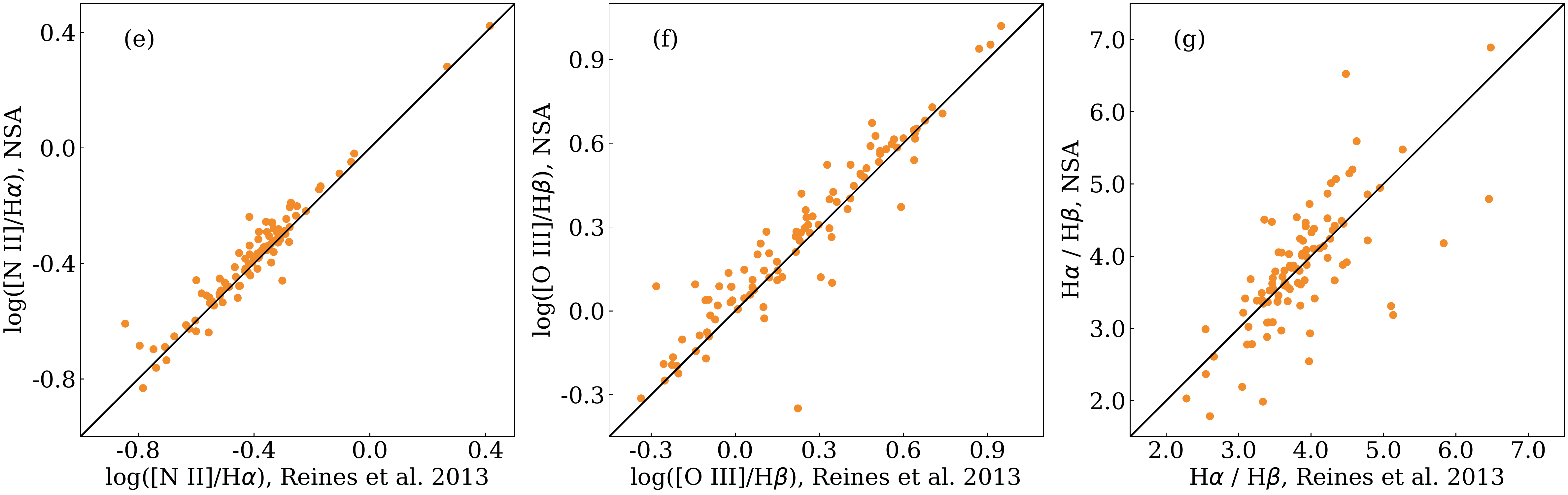}
    \caption{Comparison of the emission line measurements from the NSA and from \citet{Reines13} for the 99 dwarf galaxies without broad H$\alpha$. (a): log H$\alpha$ flux. (b): log H$\beta$ flux (c): log [N II] flux. (d): log [O III] flux (e): log [N II]/H$\alpha$. (f):  log [O III]/H$\beta$. (g) Balmer decrement, H$\alpha$/H$\beta$. All fluxes in panels (a)--(d) are in units of $10^{-17}$ erg s$^{-1}$ cm$^{-2}$. In all panels, the black lines have a slope of 1 to guide the eye.} \label{fig:ELfluxes}
\end{figure*}

\citet{Reines13} re-fit emission line fluxes for galaxies with $\log(M_*/M_{\sun}) \lesssim 9.5$ using a different procedure. To ensure uniformity when comparing our new sample with galaxies from previous surveys, the remainder of this work uses only the emission line flux measurements taken directly from the NSA. For all four relevant emission lines, the NSA has slightly higher flux measurements: the median ratios of NSA flux to \citet{Reines13} flux in the 99 dwarf galaxies are 1.07, 1.07, 1.10, and 1.16 for H$\alpha$, H$\beta$, [N {\footnotesize II}], and [O {\footnotesize III}], respectively. These are shown in panels (a) through (d) of Figure \ref{fig:ELfluxes}.

We also show a comparison of the [N {\footnotesize II}]/H$\alpha$ and [O {\footnotesize III}]/H$\beta$ emission line ratios, and the Balmer decrement, H$\alpha$/H$\beta$, using NSA and \citet{Reines13} fluxes in panels (e) through (g), as these will be used in Section \ref{sec:properties} to compare our sample of dwarf galaxies with the MCP--NSA sample. The differences between the NSA and \citet{Reines13} are less pronounced in these line ratios than for individual line measurements: the mean ratios of NSA to \citet{Reines13} emission line ratios are $1.06\pm0.13$ and $1.10\pm0.24$ for [N {\footnotesize II}]/H$\alpha$ and [O {\footnotesize III}]/H$\beta$, respectively.

As we will show in Section \ref{subsec:BPT}, using NSA fluxes has the effect of spreading out BPT classifications, such that fewer of our 99 dwarf galaxies are classified as composites, and more galaxies are classified as either star-forming or AGNs. H$\alpha$ and H$\beta$ both have identically higher mean measurements in the NSA, and the mean ratio of NSA Balmer decrement to \citet{Reines13} Balmer decrement is $1.00\pm0.16$, so the use of the NSA has, on average, no effect on our comparison.

\section{The Optical Spectroscopic Properties of Surveyed Galaxies} \label{sec:properties}

\subsection{Weighted Rate and Probability of Zero Detections} \label{subsec:weightedrate}

Based on our sample's optical properties, we calculate a weighted detection rate, $R_w$, and the corresponding weighted binomial probability of making zero detections in our sample of 100 dwarf galaxies, $P_w(0)$. To do this, we first bin both our target galaxies and the MCP--NSA sample by a host property (e.g. BPT classification, Balmer decrement), and then take the mean detection rate across these bins, weighting by the number of target dwarf galaxies in each bin. Specifically, we use the formula: 
\begin{equation} \label{equation:weightedrate}
R_w = \frac{\sum_{i} (N_{{\rm obs}, i} R_i)}{\sum_{i} N_{{\rm obs}, i}}
\end{equation}
where $N_{{\rm obs}, i}$ is the number of dwarf galaxies in our sample in the $i$th bin, and $R_i=N_{{\rm det},i}/N_{{\rm tot},i}$ is the unweighted rate in the MCP--NSA sample for the same bin. In calculating $R_w$, we only use bins where both $N_{{\rm obs}, i}$ and $N_{{\rm det}, i}$ are $\geq$ 3. Below this cut, the statistics are unreliable, potentially biasing the detection rates and artificially dominating and inflating errors on $R_w$.

We use symmetric Gaussian errors on $N_{{\rm obs}, i}$, $N_{{\rm det}, i}$, and $N_{{\rm tot}, i}$ for $N \geq 40$, and asymmetric Poisson 68\% confidence interval bounds from \citet{Gehrels86} for $N<40$. Because the Poisson errors for $N_{{\rm obs}, i}$, $N_{{\rm det}, i}$, and $N_{{\rm tot}, i}$ are asymmetric, we propagate the upper and lower errors separately, and thereby compute asymmetric errors for both $R_i$ and $R_w$. The errors on the detection rates in Figures \ref{fig:balmerdec}, \ref{fig:mstar}, \ref{fig:O3lum} and \ref{fig:O3flux} are calculated this way. 

Once we have $R_w$, we use it to calculate $P_w(0)$. We propagate the upper (lower) errors on $R_w$ to get the lower (upper) errors on $P_w(0)$, since a larger $R_w$ corresponds to a lower likelihood of making zero detections. We give $N_{{\rm det}}$, $N_{{\rm tot}}$, $R_w$ and $P_w(0)$ for each AGN property we examine in Table \ref{table:stats}. For comparison, Table \ref{table:stats} also shows the overall unweighted rate $R_0$ and unweighted probability of making zero detections out of 100 dwarf galaxies, $P_0(0)$.

\subsection{BPT Classifications} \label{subsec:BPT}
\subsubsection{AGNs, Composites, and Star-Forming Galaxies} \label{subsubsec:N2BPT}

\begin{figure*}
    \centering
    \includegraphics[width=0.88\linewidth]{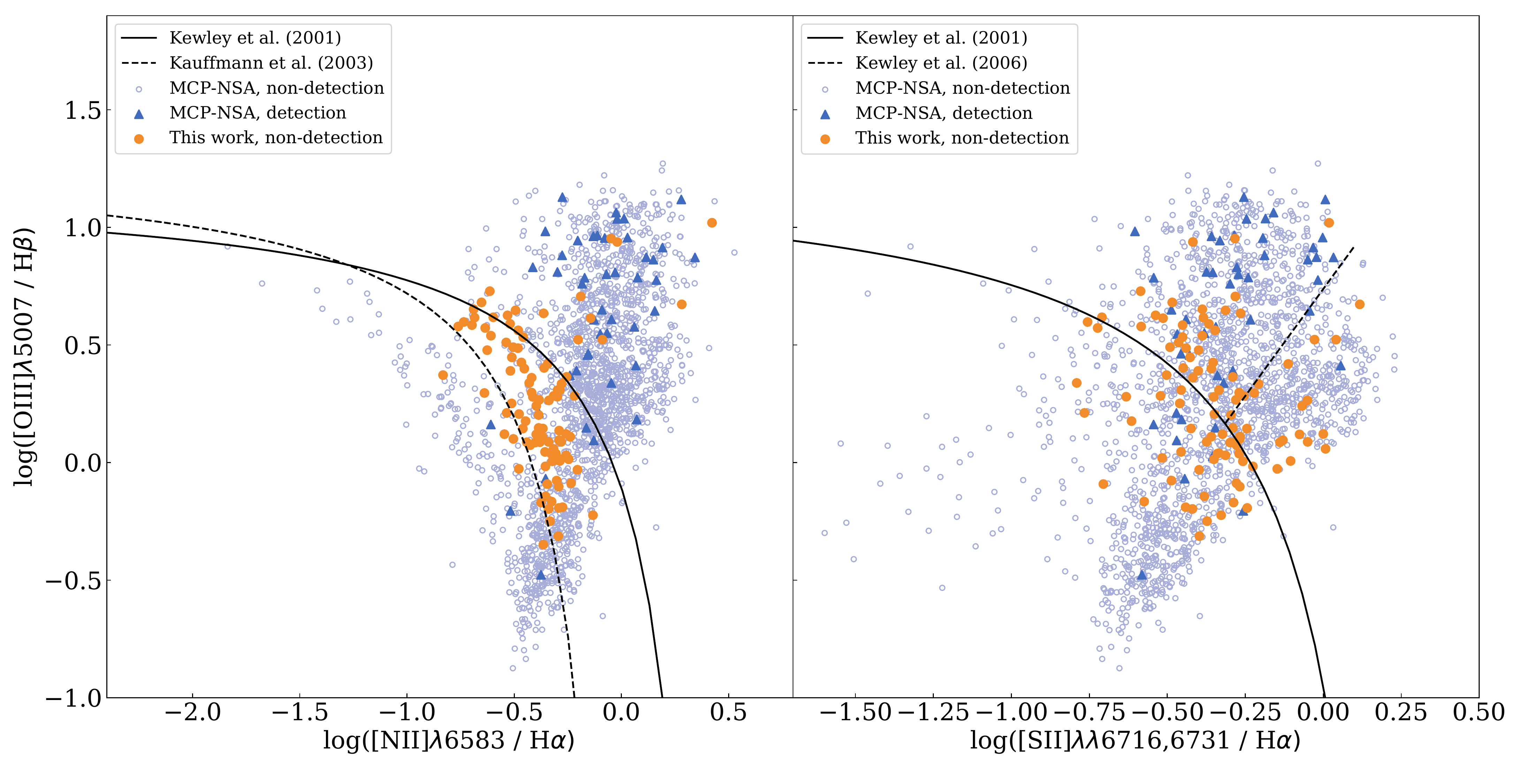}
    \caption{BPT diagrams showing the 99 dwarf galaxies in our sample as orange circles, and galaxies previously surveyed with the GBT by the MCP as blue triangles (detections) or open light-blue circles (non-detections). Emission line fluxes are taken from the NSA. Left: the BPT diagram using [N II] $\lambda6583$ / H$\alpha$ to distinguish between AGNs, composites, and star-forming galaxies, using the diagnostic criteria from \citet{Kewley01} and \citet{Kauffmann03}. Right: BPT diagram using [S II] $\lambda \lambda 6716,31$ / H$\alpha$. The dashed diagonal line denotes the demarcation between Seyferts and LINERs \citep[][Equation 6]{Kewley06}. Of the 46 MCP--NSA detections with $\rm{S/N}\geq3$ for the five emission lines, 35 are Seyfert 2 galaxies.}
    \label{fig:BPT}
\end{figure*}

Previous surveys for maser emission have targeted mostly optical AGNs \citep[e.g.,][]{Braatz04, Kondratko06a, Castangia08}, and analysis of previous maser surveys have shown that detection rate is higher in AGNs than in galaxies in general \citep[e.g.][]{Zhu11, Constantin12, Kuo18, Kuo20}. We identify AGNs among both our sample of dwarf galaxies and the MCP--NSA sample using their optical emission line ratios. We plot our targets, and both detections and non-detections from the MCP--NSA sample, on the BPT diagram for [O {\footnotesize III}] $\lambda$5007 / H$\beta$ vs. [N {\footnotesize II}] $\lambda$6583 / H$\alpha$ in Figure \ref{fig:BPT} left to distinguish between AGNs (above the \citealt{Kewley01} line), composites (between the \citealt{Kewley01} and \citealt{Kauffmann03} lines), and star-forming galaxies (below the \citealt{Kauffmann03} line).

%	 BPT Class   N_DET_NSA N_NON_NSA N_TOT_NSA RATE_NSA N_NON_DMS P_BINOM_0_DMS
%	------------ --------- --------- --------- -------- --------- -------------
%	         AGN        38      1058      1096   3.4672        17        3.0396
%	   Composite         5       397       402   1.2438        74       28.9649
%	Star-forming         3       346       349   0.8596         8       42.5418
%
%	For 99 objects...
%		R_0 = 46/1847 = 2.49%
%		R_w = 1.59%
%		P_0(0) = 8.23%
%		P_w(0) = 20.37%

To consistently compare galaxies from our target sample with those from the MCP--NSA sample, we exclusively use emission line flux measurements from the NSA, which differ slightly from those measured by \citet{Reines13}. Whereas \citet{Reines13} identified 88 composites and 11 AGNs, with measurements from the NSA for the same sample we identify 8 star-forming galaxies, 74 composites, and 17 AGNs. The majority of galaxies in the MCP--NSA sample are AGNs, for which the detection rate is significantly higher ($3.5\pm0.6\%$) than for composites ($1.2^{+0.9}_{-0.5}\%$) and star-forming galaxies ($0.9^{+0.8}_{-0.5}\%$).

Only 17\% of our sample are AGNs, whereas most of the MCP--NSA sample falls above the \citet{Kewley01} line (1,096, or 59\%, of 1,847 galaxies in the sample). The [N {\footnotesize II}] / H$\alpha$-weighted detection rate is $R_w=1.59^{+0.73}_{-0.53}\%$, compared to the unweighted rate of $R_0=2.49\pm0.37\%$ (46/1847). This corresponds to a weighted probability of no detections in our sample of 100 dwarf galaxies of $P_w(0) = 20.04^{+10.86}_{-14.95}\%$, as opposed to the unweighted probability of $P_0(0) = 8.03\pm3.06\%$.

\subsubsection{Seyferts and LINERs} \label{subsubsec:S2BPT}

%	 BPT Class   N_DET_NSA N_NON_NSA N_TOT_NSA RATE_NSA N_NON_DMS P_BINOM_0_DMS
%	------------ --------- --------- --------- -------- --------- -------------
%	   Ambiguous         1        69        70   1.4286         3       24.0625
%	   Composite         5       394       399   1.2531        74       28.6961
%	       LINER         2       371       373   0.5362         3       58.7273
%	     Seyfert        35       620       655   5.3435        14        0.4354
%	Star-forming         3       335       338   0.8876         5       41.3686
%
%	For 99 objects...
%		R_0 = 46/1835 = 2.50%
%		R_w = 1.80%
%		P_0(0) = 8.10%
%		P_w(0) = 16.61%

In previous surveys, the detection rate has been higher in Seyferts than in AGNs in general \citep[e.g.][]{Braatz97, Zhu11, Constantin12}. We therefore subdivide the AGNs into Seyferts and LINERs. \citet{Kewley06} distinguish between Seyferts and LINERs using [S {\footnotesize II}] $\lambda\lambda$6716,6731 / H$\alpha$ and [O {\footnotesize I}] $\lambda$6300 / H$\alpha$ emission line ratios, but of the 99 galaxies we observe that are in the NSA, 40 have [O {\footnotesize I}] line measurements with $\rm{S/N}<2$. Therefore, we use only [S {\footnotesize II}] / H$\alpha$.

We plot all our targets, and both the detections and non-detections in the MCP--NSA sample, on the BPT diagram for [O {\footnotesize III}] / H$\beta$ vs. [S {\footnotesize II}] / H$\alpha$ in Figure \ref{fig:BPT} right. The solid black curves in the left and right panels show the AGN/star-forming demarcations given by Equations 5 and 6 from \citet{Kewley01}, respectively. For a galaxy to be classified as either a Seyfert or LINER, it must lie above \textit{both} of these curves. Seyferts and LINERs are then separated by the dashed line in the right panel of Figure \ref{fig:BPT}, which shows Equation 6 from \citet{Kewley06}. [N {\footnotesize II}] composites remain classified as composites, and galaxies that lie below both the dashed curve from \citet{Kauffmann03} on Figure \ref{fig:BPT} left, and below the solid curve of \citet{Kewley01} on Figure \ref{fig:BPT} right, are classified as star-forming galaxies.

There are a handful of galaxies in the MCP--NSA sample, and three galaxies among our targets, for which the [N {\footnotesize II}] and [S {\footnotesize II}] BPT classifications disagree, i.e. one suggests ionization related to star-formation and the other suggests ionization by an AGN. We classify these galaxies as ``Ambiguous.''

Classifying all 99 galaxies that we have observed from the NSA using the above criteria, there are 14 Seyferts, 3 LINERs, 74 composites, 5 star-forming galaxies, and 3 galaxies with Ambiguous classifications. We find that Seyferts comprise the plurality of archival observations, with 36\% of MCP--NSA galaxies with good [S {\footnotesize II}] S/N being Seyferts. The detection rate in MCP--NSA galaxies is highest in Seyferts, at $\sim$5\%, and Seyferts comprise 35 of 46 MCP--NSA detections. Previous studies quote the detection fraction in Seyferts being as high as $\sim$7--8\% \citep[e.g.][]{Braatz97, Zhu11, Constantin12}, though we note that some of these authors used the demarcation lines from \citet{VO87}: $[\rm{N~II}]/\rm{H}\alpha=0.6$ and $[\rm{O~III}]/\rm{H}\beta=3$, with Seyferts (LINERs) lying above (below) the latter line.

Though the emission line ratios of Seyferts are bona fide indicators of accretion onto an SMBH, some studies have suggested that LINER-like nuclei, especially those with lower luminosities, could be ionized by sources other than accretion onto a massive BH \citep[e.g.][]{Ho08, CidFernandes11, YanBlanton12}. In addition to the 3 known LINERs, nearly half of the composites (34 out of 74) have LINER-like line ratios that would fall below an extended version of the \citet{Kewley06} line in Figure \ref{fig:BPT} right, and the majority (62 out of 74) fall below the $[\rm{O~III}]/\rm{H}\beta=3$ line from \citet{VO87}. As we will explore further in Section \ref{subsec:lumandmass}, the dwarf galaxy sample spans a low range in [O {\footnotesize III}] luminosity, a proxy for bolometric luminosity, compared to the massive galaxies in other maser surveys. If some of these low-luminosity LINERs and LINER-like composites indeed do not host true AGNs, it could be a significant contributing factor to our lack of detections.

There are too few MCP detections with LINER and Ambiguous classifications to use as separate categories when calculating $R_w$. To account for them, we combine these galaxies with the composites. Together, they comprise galaxies with emission line ratios where the underlying ionization mechanism is mixed or unclear. These galaxies have a detection rate of only $\sim$1\%. We find $R_w=1.57^{+0.56}_{-0.45}\%$, corresponding to $P_w(0)=20.58^{+9.41}_{-11.68}\%$.

\subsection{Obscuration and Geometric Effects} \label{subsec:obscuration}

\begin{figure}
    \centering
    \includegraphics[width=0.95\linewidth]{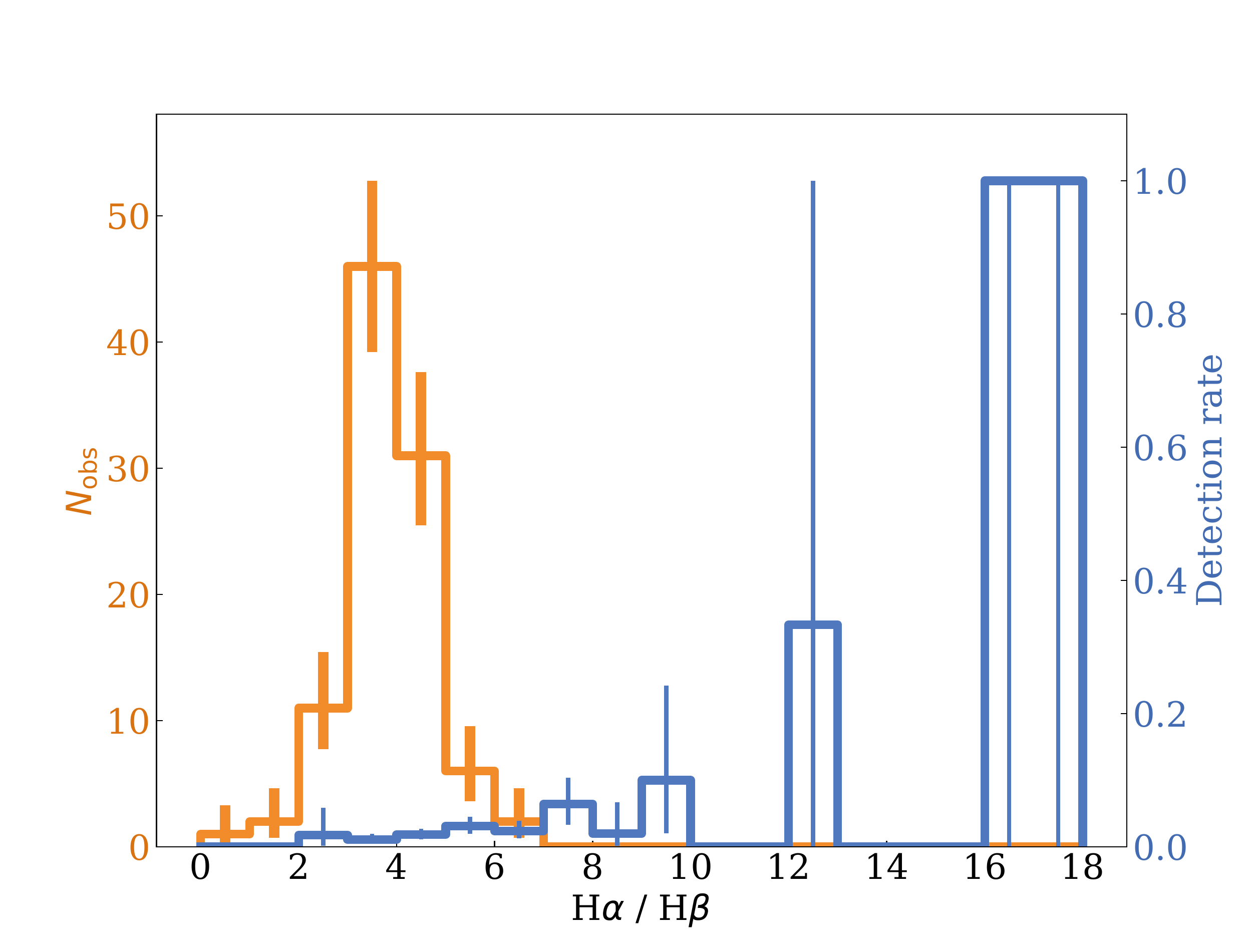}
    \caption{A comparison of our sample against detection rates in previous surveys with respect to the observed Balmer decrement. The orange histogram shows the number of galaxies in our survey sample in each bin of $\rm{H}\alpha/\rm{H}\beta$, all of which are non-detections. The blue histogram shows the detection rate $N_{{\rm det}}/N_{{\rm tot}}$ from the MCP--NSA sample in each bin. The errors for both the orange and blue histograms are computed using Gaussian errors for $N\geq40$ and Poisson errors for $N<40$. Data were binned by 1 in Balmer decrement. All H$\alpha$ and H$\beta$ fluxes are taken directly from the NSA.}
\label{fig:balmerdec}
\end{figure}

Maser emission requires long amplification gain paths along the line of sight. Consequently, the maser detection fraction is higher in highly-obscured nuclei \citep[e.g.][]{Zhang06, Greenhill08, Zhang10, Castangia19}. In optical spectroscopy, this geometric effect is marked by a higher detection fraction in galaxies with higher Balmer decrement $\rm{H}\alpha/\rm{H}\beta$ \citep[e.g.][]{Zhu11}. We compare the Balmer decrement of the dwarf galaxies in our target sample to the MCP--NSA galaxies, and find that they are much less obscured compared to the massive galaxies from pervious surveys. The mean Balmer decrement of the dwarf galaxies in our survey is 3.81, whereas the mean Balmer decrement is 5.11 in the 1,801 non-detections in the MCP--NSA, and 6.34 in the 46 detections.

The detection rate in the archival MCP--NSA sample and the number of dwarf galaxies in our observations are plotted as functions of Balmer decrement Figure \ref{fig:balmerdec}. As in Section \ref{subsec:BPT}, we can use Equation \ref{equation:weightedrate}, but summing over bins in Balmer decrement instead of BPT classifications. To increase the number of bins with $N_{{\rm det}}$ and $N_{{\rm obs}}$ $\geq$ 3, we compute $R_w$ using a bin width of 2. The $\rm{H}\alpha/\rm{H}\beta$-weighted detection fraction is $R_w=1.56^{+0.61}_{-0.47}\%$, corresponding to $P_w(0)=20.80^{+9.92}_{-12.84}\%$.

There are two likely explanations for why our sample is under-obscured compared to the MCP--NSA galaxies, despite having selected only Type 2 AGNs as targets. First, since we performed this selection by removing galaxies in which \citet{Reines13} detected broad H$\alpha$ emission, our sample may include unobscured AGNs where broad Balmer emission is simply too faint to detect, as might be the case for the BLR around an IMBH. Indeed, \citet{Bianchi12} confirm the existence of three unabsorbed Seyfert 2 AGNs with low accretion rates. Such low-luminosity, unobscured Type 2 AGNs are less likely to meet the geometric requirements for maser emission.

Second, more obscured AGNs in dwarfs may be harder to detect spectroscopically. H$\beta$ is the weakest of the four lines used for spectroscopic identification of AGNs, and can be lost in highly-obscured systems. \citet{Reines13} pre-selected for galaxies with S/N(H$\beta)\geq2$, and obscured AGNs in dwarf galaxies may have been removed from the sample by this cut.

\subsection{Maser Luminosity and Host AGN Properties} \label{subsec:lumandmass}

In the model of disc maser emission by \citet{NeufeldMaloney95}, the critical radius within which the physical conditions for maser emission are met scales with both AGN luminosity and BH mass: $R_{\rm{cr}} \propto M_{\rm{BH}}^{0.62}L_{\rm{AGN}}^{0.38}$. Assuming that maser luminosity correlates with $R_{\rm{cr}}$, it should also correlate with both $M_{\rm{BH}}$ and $L_{\rm{AGN}}$.

Accurate measurements of $M_{\rm{BH}}$ and $L_{\rm{AGN}}$ are not available for most galaxies, and analyses must rely on proxies for these two properties. Previous studies have taken advantage of the tight correlation between $M_{\rm{BH}}$ and $\sigma_*$ \citep[e.g.][]{FerrareseMerritt2000, Gultekin09, KormendyHo13, Greene19}, and used $\sigma_*$ as a proxy for $M_{\rm{BH}}$. These studies have confirmed that the maser detection rate increases with increasing $\sigma_*$ \citep[e.g.][]{Zhu11, vandenBosch16}. However, this is not possible for our sample, as only 15 of the dwarf galaxies in our survey have well-measured $\sigma_*$ (i.e., not flagged, with positive S/N, and above the SDSS spectral resolution limit of $\sim$70 km s$^{-1}$).

Instead, in Section \ref{subsubsec:mass}, we inspect the detection rate with respect to the total galaxy stellar mass $M_*$. Stellar masses are provided in the NSA, and are known to correlate with $M_{\rm{BH}}$, though the relation is less tight than for $\sigma_*$ \citep[e.g.][]{ReinesVolonteri15, Greene19}. Additionally, it is important to analyse the effect of $M_*$ on the detection rate in previous surveys, since our sample is defined by a cut of $M_* \lesssim 10^{9.5}~M_{\sun}$.

In addition to $M_{\rm{BH}}$, maser luminosity should scale with $L_{\rm{AGN}}$, for which the luminosity of the [O {\footnotesize III}] $\lambda$5007 emission line is a known proxy \citep[e.g.][]{Heckman05, HeckmanBest14}. The strength of the [O {\footnotesize III}] line has been shown to correlate with the maser detection rate in massive galaxies \citep[e.g.][]{Zhu11, Constantin12, vandenBosch16}. We compare our sample's distribution in $L_{\rm{[O~III]}}$ and the detection rate as a function of $L_{\rm{[O~III]}}$ in Section \ref{subsubsec:luminosity}.

Accurate distances are required to calculate both masses and luminosities. We determine distances using the ZDIST parameter from the NSA, which gives galaxy redshift corrected for its peculiar velocity in the Local Group frame of reference \citep{Willick97}. Distance is calculated as $D= {\rm ZDIST} \times (c/H_0)$. To ensure the reliability of this conversion given some NSA galaxies are sufficiently nearby that their peculiar velocities could be significant ($D$ = 3--15 Mpc), in the remainder of this paper, we limit our analysis to galaxies with ZDIST/ZDIST\_ERR $>$ 10. This cut removes four detections and 24 non-detections from the MCP--NSA sample, and five of the 99 dwarf galaxies from our target sample.

\subsubsection{Stellar Mass} \label{subsubsec:mass}

\begin{figure}
    \centering
    \includegraphics[width=0.95\linewidth]{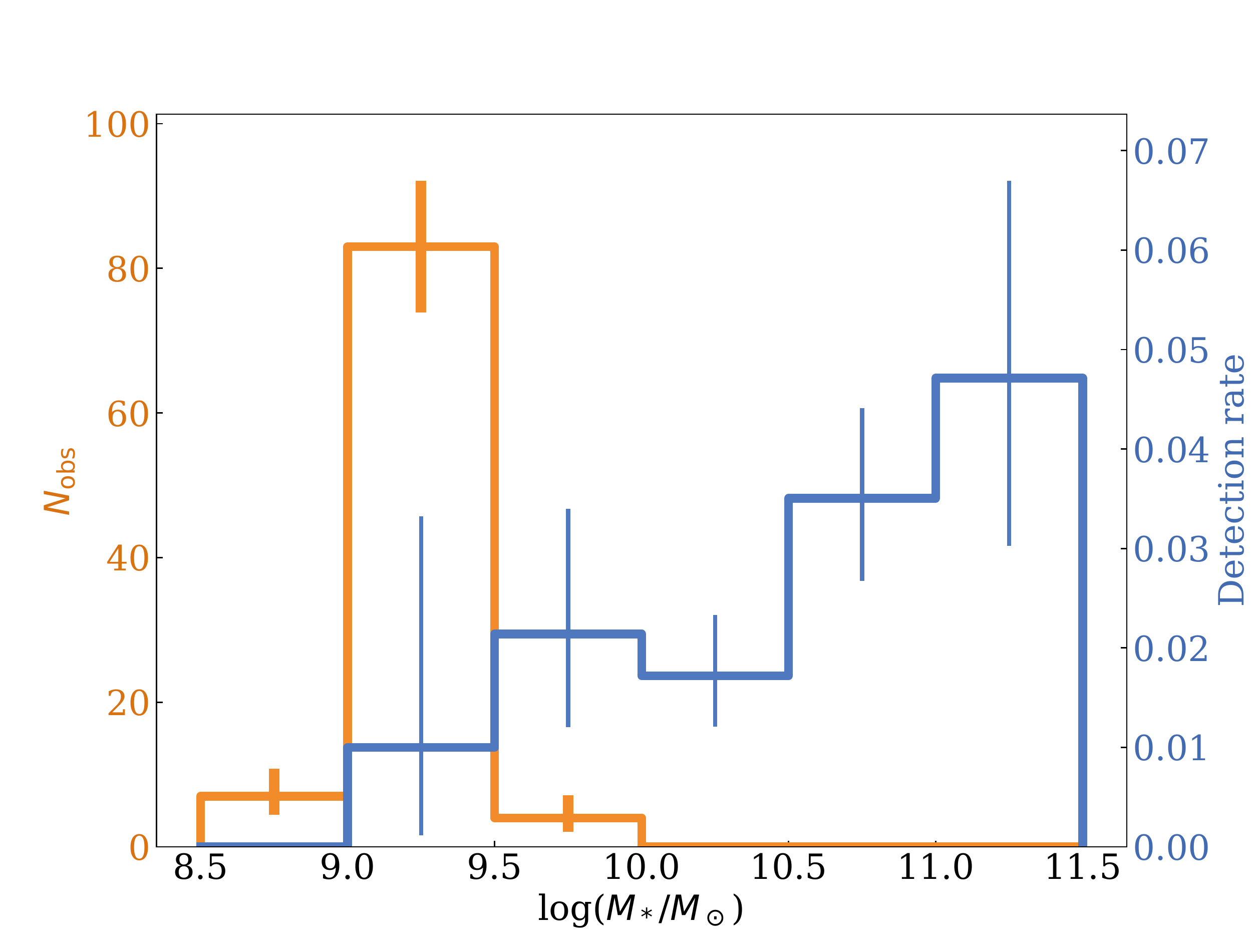}
    \caption{A comparison of our sample against detection rates in previous surveys with respect to total galaxy stellar mass. The orange histogram shows the number of galaxies in our survey sample in each bin of $M_*$, all of which are non-detections. The blue histogram shows the detection rate $N_{{\rm det}}/N_{{\rm tot}}$ from previous surveys in each bin. In this histogram, we include matched galaxies with any spectroscopic S/N, provided they have good ZDIST and good velocity agreement, as $M_*$ is not calculated from spectroscopic measurements. The errors for both the orange and blue histograms are computed using Gaussian errors for $N\geq40$ and Poisson errors for $N<40$. Data were binned by 0.5 dex in $M_*$. Stellar masses are taken from the NSA, which uses the \texttt{kcorrect} code from \citet{BlantonRoweis07}, and were calculated using $h=0.70$. There was a single GBT detection and two non-detections in the $M_* = 10^{12-12.5}~M_{\sun}$ bin. This plot is therefore truncated to $M_*\leq10^{11.5}~M_{\sun}$, to show the trend in detection rate in bins with higher counts.}
\label{fig:mstar}
\end{figure}

The stellar masses for the dwarf galaxies in our sample are taken directly from the NSA, which uses the \texttt{kcorrect} code from \citet{BlantonRoweis07}. Given that these $M_*$ measurements are available for all NSA galaxies and do not depend on spectroscopic measurements, we compare $M_*$ for all matched galaxies with good velocity agreement and good ZDIST. Of the 2,975 galaxies with good velocity agreement, there are 78 GBT detections and 2,877 total galaxies that also have good ZDIST, for an unweighted detection rate of $R_0=2.71\pm0.31\%$.

Stellar masses are inversely proportional to the square of the dimensionless Hubble parameter, $h$. \citet{Reines13} cut at $M_*\leq10^{9.5}~M_{\sun}$ for masses calculated using $h=0.73$. Recalculating $M_*$ for $h=0.70$ increases all stellar masses by a factor of $(0.73/0.70)^2=8\%$, and the maximum mass among our target dwarf galaxies after applying this increase is $10^{9.51}~M_{\sun}$. The mean stellar mass in our sample of dwarf galaxies is $10^{9.31}~M_{\sun}$, while the mean stellar mass of the detections (non-detections) in this good-ZDIST sample is $10^{10.65}~M_{\sun}$ ($10^{10.41}~M_{\sun}$). We plot a histogram of the detection rate in previous surveys distribution of our dwarf galaxy sample with respect to $M_*$ in Figure \ref{fig:mstar}. The $\log(M_*/M_{\sun})=$ 12.0--12.5 bin has only a single MCP detection and two non-detections. We therefore do not include it in the histogram, in order to show the trend in the bins with higher counts.

Figure \ref{fig:mstar} shows that the detection rate increases weakly with $M_*$. Our sample lies primarily in bins with lower detection rates. The trend in detection rate is slightly less pronounced than the trend for the other host properties we have analysed. This may reflect that $M_*$ is a looser proxy for $M_{\rm{BH}}$ than, for example, $\sigma_*$, which has been shown in previous analyses to relate more closely with the detection rate \citep[e.g.][]{Zhu11}.

There are 496 low-mass galaxies ($M_*<10^{10}~M_{\sun}$) in this sample, of which 169 are dwarf galaxies. Eight of the low-mass galaxies are known masers, giving a detection rate of 1.6$^{+0.8}_{-0.6}$\%. Only one of the known masers is in a dwarf galaxy with a mass of $M_*=10^{9.31}~M_{\sun}$, and the remaining seven have masses between $10^{9.5}~M_{\sun}$ and $10^{10}~M_{\sun}$. We give the properties of the low-mass galaxies hosting maser emission and show their maser spectra in Appendix \ref{sec:low-mass-masers}. Since most of the targets in previous maser surveys are massive galaxies, there is little overlap between the range of $M_*$ for our sample of dwarf galaxies and that for the archival sample; only a single bin, with $N_{\rm{obs}}=4$ and $N_{\rm{det}}=7$, has enough detections and observed dwarf galaxies to be used to calculate $R_w$. Therefore, we do not report an $M_*$-weighted detection rate for our surveyed sample.

\subsubsection{[O III] Luminosity} \label{subsubsec:luminosity}

\begin{figure}
    \centering
    \includegraphics[width=0.95\linewidth]{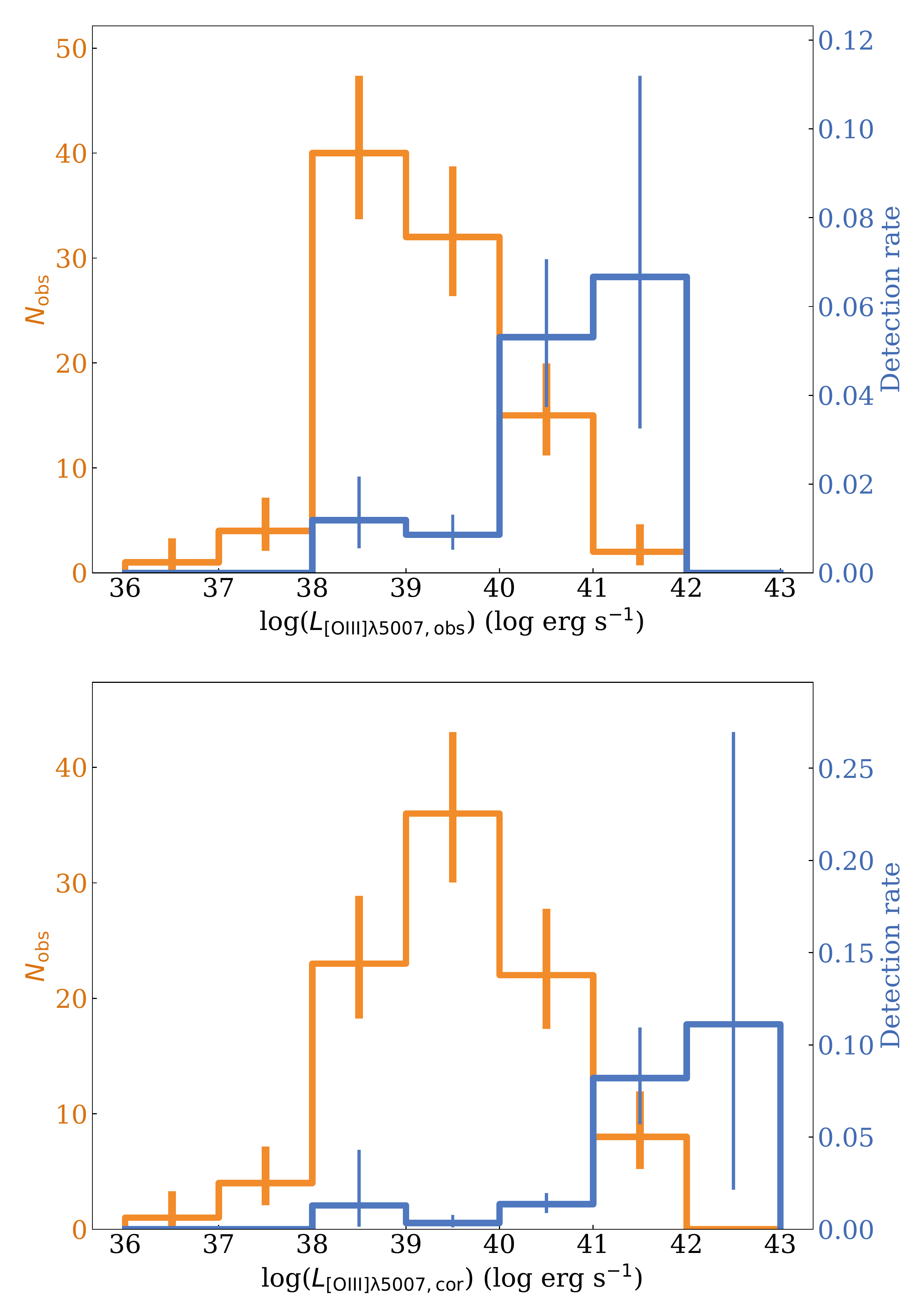}
    \caption{A comparison of our sample against detection rates in previous surveys with respect to [O III] $\lambda5007$ luminosity. The orange histogram shows the number of galaxies we surveyed per bin. The blue histogram shows the detection rate $N_{{\rm det}}/N_{{\rm tot}}$ from the MCP--NSA sample in each bin. Top: observed [O III] luminosity. Bottom: extinction-corrected [O III] luminosity, using the correction formula from \citet{Bassani99} and an assumed intrinsic Balmer decrement $(\rm{H}\alpha/\rm{H}\beta)_0=3$.  The errors for both the orange and blue histograms are computed using Gaussian errors for $N\geq40$ and Poisson errors for $N<40$. Data were binned by 1.0 dex in [O III] luminosity. All [O III] fluxes were taken directly from the NSA.}
\label{fig:O3lum}
\end{figure}

The mean observed [O {\footnotesize III}] luminosity for the dwarf galaxies in our sample is $1.5\times10^{39}$ erg s$^{-1}$. The mean value for the MCP--NSA non-detections of $4.4\times10^{39}$ erg s$^{-1}$, and the mean value for the 36 MCP--NSA detections is $1.8\times10^{40}$ erg s$^{-1}$, more than an order of magnitude greater than that of our sample. We show the detection rate among the matched MCP--NSA sample and the distribution of our targets against $L_{\rm{[O~III}],obs}$ in the top panel of Figure \ref{fig:O3lum}.

Because the central engine is typically highly obscured in the optical, but unobscured at radio wavelengths, maser luminosities should be more closely related to the intrinsic [O {\footnotesize III}] luminosity, corrected for optical extinction. We use the correction formula from \citet{Bassani99}, $L_{\rm{[O~III]},corr} = L_{\rm{[O~III],obs}} \times [(\rm{H}\alpha / \rm{H}\beta)/(\rm{H}\alpha / \rm{H}\beta)_0]^{2.94}$ where the intrinsic Balmer decrement $(\rm{H}\alpha / \rm{H}\beta)_0$ is assumed to be 3.

As mentioned in Section \ref{subsec:obscuration}, our sample of dwarf galaxies is under-obscured relative to previously surveyed galaxies. Consequently, the discrepancy between our sample and the MCP--NSA sample is even more pronounced in extinction-corrected luminosity: The mean value of $L_{\rm{[O~III]},corr}$ is $3.3\times10^{39}$ erg s$^{-1}$ among our dwarf galaxy targets, $1.8\times10^{40}$ erg s$^{-1}$ for the non-detections in the MCP--NSA sample, and $1.3\times10^{41}$ erg s$^{-1}$ for MCP--NSA detections. The weighted rate for $L_{\rm{[O~III]},obs}$ is $R_w=1.65^{+0.63}_{-0.48}\%$, while the weighted rate for $L_{\rm{[O~III]},corr}$ is $R_w=1.02^{+0.46}_{-0.36}\%$. These correspond to weighted probabilities of making zero detections of $P_w(0)=19.03^{+9.35}_{-12.16}\%$ and $36.01^{+13.23}_{-16.72}\%$ when weighting by observed and corrected luminosity, respectively.

\subsubsection{[O III] Flux} \label{subsubsec:flux}

\begin{figure}
    \centering
    \includegraphics[width=0.95\linewidth]{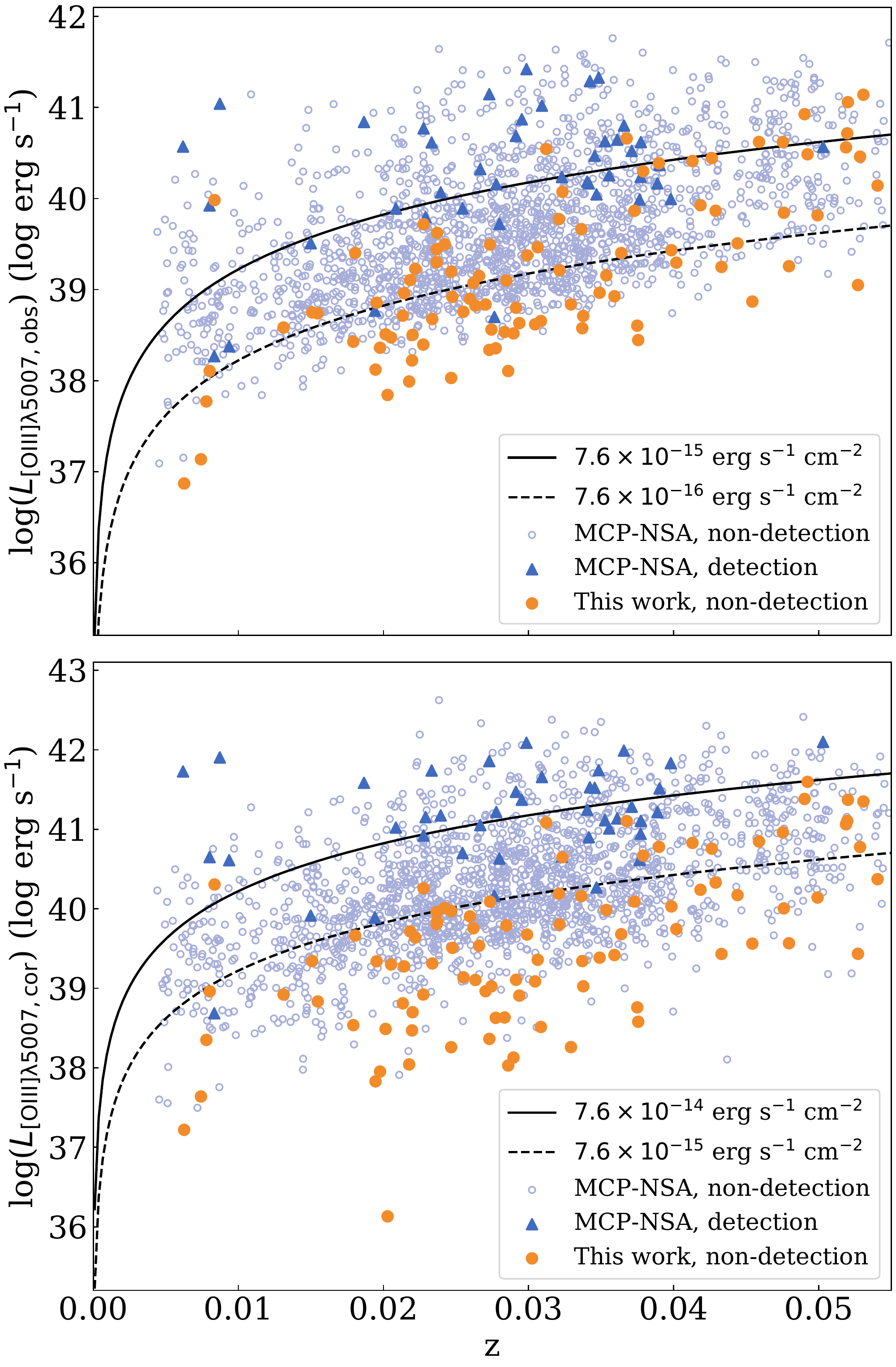}
    \caption{Observed and corrected [O III] luminosity vs redshift. Redshifts shown use the ZDIST parameter from the NSA, which accounts for the peculiar velocities of the galaxies \citep{Willick97}. We include only galaxies with accurate luminosity distances, i.e. $\rm{ZDIST/ZDIST\_ERR}>10$. Most galaxies with $z\lesssim0.005$ are removed via this cutoff. Top: observed [O III] luminosity vs redshift. Bottom: extinction-corrected [O III] luminosity vs redshift, using the correction formula from \citet{Bassani99}. For both observed and corrected [O III] flux, the solid lines show the [O III] flux that corresponds to the H$_2$O flux limit of 0.1 Jy km s$^{-1}$, from \citet{Zhu11}. The dashed shows the same flux limit, but shifted --1.0 dex, to account for the large scatter in the correlation between [O III] and maser emission. In both panels, orange circles show the non-detections in dwarf galaxies, surveyed in this work, while blue triangles (light blue, open circles) show the detections (non-detections) from the MCP--NSA sample.}
\label{fig:O3vsz}
\end{figure}

\begin{figure}
    \centering
    \includegraphics[width=0.95\linewidth]{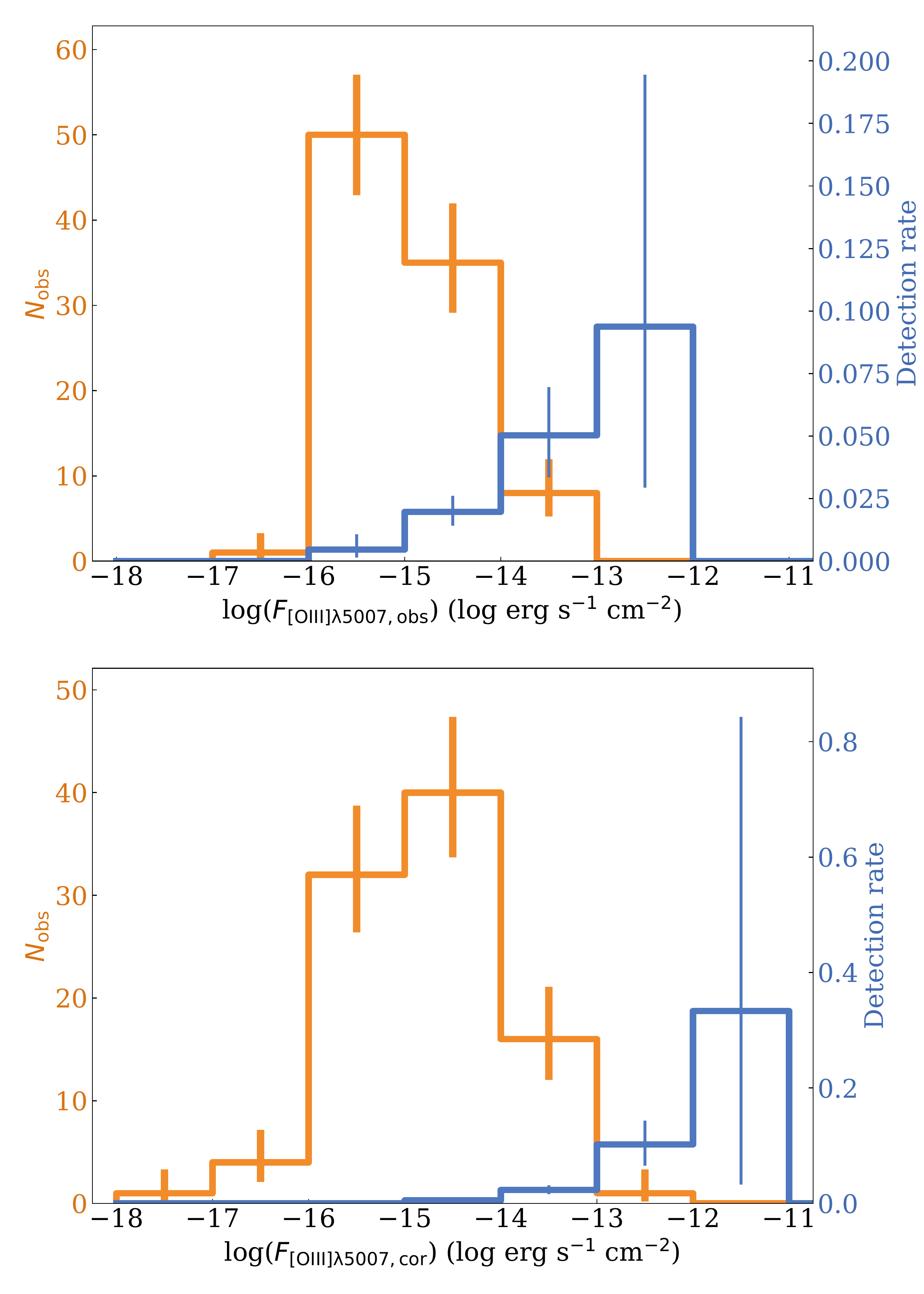}
    \caption{Similar to Figure \ref{fig:O3lum}, but with flux instead of luminosity. The orange histogram shows the number of galaxies in our survey sample in each bin of [O III] flux. The blue histogram shows the detection rate $N_{\rm{det}}/N_{\rm{tot}}$ from the MCP--NSA sample in each bin. Top: observed [O III] flux. Bottom: [O III] flux corrected for extinction, using the correction formula from \citet{Bassani99} and an assumed intrinsic Balmer decrement $(\rm{H}\alpha/\rm{H}\beta)_0=3$. The errors for both the orange and blue histograms are computed assuming Gaussian errors for $N\geq40$ and Poisson errors for $N<40$. Data were binned by 1.0 dex in [O III] flux. All emission line fluxes were taken directly from the NSA.}
\label{fig:O3flux}
\end{figure}

Our survey uses equal observation time with an identical setup for each galaxy and is, therefore flux limited. Surveys for masers in mostly massive galaxies have been shown to also be flux limited. \citet{Zhu11} show that masers tend to have integrated fluxes above a limit of $\sim$0.1 Jy km s$^{-1}.$\footnote{This is computed assuming a 3$\sigma$ flux density limit of 10 mJy, over a complex of blended Doppler components spanning 10 km s$^{-1}$, though see Figure 9 and Section 4.2.1. of \citet{Kuo20} for alternative flux limits.} Therefore, we inspect the [O {\footnotesize III}] flux of galaxies in our survey and the MCP--NSA sample, rather than luminosity, as a proxy for expected H$_2$O flux.

\citet{Zhu11}, based on an observed correlation between [O {\footnotesize III}] flux and maser flux, propose [O {\footnotesize III}] flux limits above which maser detection is more efficient, namely $7.6 \times 10^{-15}$ erg s$^{-1}$ cm$^{-2}$ in observed [O {\footnotesize III}] flux and $7.6 \times 10^{-14}$ erg s$^{-1}$ cm$^{-2}$ in extinction-corrected [O {\footnotesize III}] flux. Due to the large scatter, they also show these limits shifted by --1.0 dex. We plot $L_{\rm{[O~III]}}$ vs redshift for the 94 dwarf galaxies with good ZDIST in our survey, as well as both detections and non-detections in the MCP, and compare them against the flux limits, which are shown as solid and dashed lines. For both observed and extinction-corrected [O {\footnotesize III}], the majority of dwarf galaxies in our target sample fall below even the downshifted flux limit. 

\begin{table*}
    \centering
    \caption{Statistics of the survey taking host galaxy properties into account}  \label{table:stats}
    \begin{tabular}{@{\extracolsep{8pt}}lccccccc@{}}
        \hline \hline 
        Galaxy property & \multicolumn{3}{c}{MCP--NSA} & \multicolumn{4}{c}{Newly--surveyed galaxies} \\
        \cline{2-4} \cline{5-8}
         & $N_{\rm{det}}$ & $N_{\rm{tot}}$ & $R_0$ & $N_{\rm{obs}}$ & $R_w$ & $P_0(0)$ & $P_w(0)$ \\
        (1) & (2) & (3) & (4) & (5) & (6) & (7) & (8) \\
        \hline
        No weighting, good velocity agreement & 89 & 2975 & $2.99\pm0.32$ & 99 & -- & $4.80\pm1.59$ & -- \\
        \hline
        [N {\footnotesize II}] / H$\alpha$ Classification: & & & & & & & \\
        \vspace{3pt}
        AGN & 38 & 1096 & $3.47^{+0.89}_{-0.82}$ & 17 & -- & -- & -- \\
        \vspace{3pt}
        Composite & 5 & 402 & $1.24^{+0.89}_{-0.60}$ & 74 & -- & -- & -- \\
        \vspace{3pt}
        Star-forming & 3 & 349 & $0.86^{+0.86}_{-0.51}$ & 8 & -- & -- & -- \\
        Total  & 46 & 1847 & $2.49\pm0.37$ & 99 & $1.59^{+0.73}_{-0.53}$ & $8.03\pm3.06$ & $20.04^{+10.86}_{-14.95}$ \\
        \hline
        [S {\footnotesize II}] / H$\alpha$ + [N {\footnotesize II}] / H$\alpha$ Classification: & & & & & & & \\
%        Seyfert & 35 & 655 & 5.34 & 14 & -- & -- & -- \\ 
%	LINER & 2 & 373 & 0.54 & 3 & -- & -- & -- \\
%        Composite & 5 & 399 & 1.25 & 74 & -- & -- & -- \\
%	Star-forming & 3 & 338 & 0.89 & 5 & -- & -- & -- \\
%	Ambiguous & 1 & 70 & 1.43 & 3 & -- & -- & -- \\
%        Total & 46 & 1835 & $2.51\pm0.37$ & 99 & 1.80 & $8.10\pm3.04$ & 16.61 \\
        \vspace{3pt}
        Seyfert & 35 & 655 & $5.34^{+1.50}_{-1.39}$ & 14 & -- & -- & -- \\ 
        \vspace{3pt}
        Composite + LINER + Ambiguous & 8 & 842 & $0.95^{+0.50}_{-0.37}$ & 80 & -- & -- & -- \\
        \vspace{3pt}
	Star-forming & 3 & 338 & $0.89^{+0.89}_{-0.53}$ & 5 & -- & -- & -- \\
        Total & 46 & 1835 & $2.51\pm0.37$ & 99 & $1.57^{+0.56}_{-0.45}$ & $7.90\pm3.03$ & $20.58^{+9.41}_{-11.68}$ \\
        \hline
        Balmer decrement & 46 & 1847 & $2.49\pm0.37$ & 99 & $1.56^{+0.61}_{-0.47}$ & $8.03\pm3.06$ & $20.80^{+9.92}_{-12.84}$ \\
        \hline
        \vspace{3pt}
        $L_{\rm{[O~III],obs}}$ & 42 & 1818 & $2.31\pm0.36$ & 94 & $1.65^{+0.63}_{-0.48}$ & $9.66\pm3.56$ & $19.03^{+9.35}_{-12.16}$ \\
        $L_{\rm{[O~III],corr}}$ & 42 & 1818 & $2.31\pm0.36$ & 94 & $1.02^{+0.46}_{-0.36}$ & $9.66\pm3.56$ & $36.01^{+13.23}_{-16.72}$ \\
        \hline
        \vspace{3pt}
        $F_{\rm{[O~III],obs}}$ (good ZDIST) & 42 & 1818 & $2.31\pm0.36$ & 94 & $1.16^{+0.41}_{-0.34}$ & $9.66\pm3.56$ & $31.08^{+10.68}_{-12.81}$ \\
        \vspace{3pt}
        $F_{\rm{[O~III],corr}}$ (good ZDIST) & 42 & 1818 & $2.31\pm0.36$ & 94 & $0.61^{+0.27}_{-0.20}$ & $9.66\pm3.56$ & $54.21^{+11.05}_{-14.54}$ \\
        \vspace{3pt}
        $F_{\rm{[O~III],obs}}$ (any ZDIST) & 46 & 1847 & $2.49\pm0.37$ & 99 & $1.57^{+0.56}_{-0.42}$ & $8.03\pm3.06$ & $20.50^{+8.79}_{-11.73}$ \\
        $F_{\rm{[O~III],corr}}$ (any ZDIST) & 46 & 1847 & $2.49\pm0.37$ & 99 & $0.65^{+0.27}_{-0.21}$ & $8.03\pm3.06$ & $52.15^{+11.01}_{-14.34}$ \\
        \hline
    \end{tabular}
    \begin{tablenotes}
        \item (1) --- The galaxy property/selection criteria by which the detection rate is weighted.
        \item (2) --- Number of detections in the MCP--NSA sample for the given property/selection criteria. For the total under [S II] / H$\alpha$, this is limited to galaxies with $\rm{S/N_{[S~II]}} > 3$, in addition to the other spectroscopic S/N requirements. For [O III] luminosity, this is limited to galaxies with ZDIST/ZDIST\_ERR $>10$.
        \item (3) --- Total number of galaxies (detections and non-detections) in the MCP--NSA sample for the given property/selection criteria. For the total under [S II] / H$\alpha$, this is limited to galaxies with $\rm{S/N_{[S~II]}} > 3$, in addition to the other spectroscopic S/N requirements. For [O III] luminosity, this is limited to galaxies with ZDIST/ZDIST\_ERR $>10$. [O III] flux is shown both with and without the ZDIST requirement.
        \item (4) --- The unweighted detection rate, expressed in per cent.
        \item (5) --- Number of dwarf galaxies in our new GBT sample for the given property/selection criteria. For [O III] luminosity, this is limited to galaxies with ZDIST/ZDIST\_ERR $>10$, of which there are 94. [O III] flux is shown both with and without the ZDIST requirement.
        \item (6) --- The detection rate, weighted by the distribution of the given property/selection criteria for our surveyed sample, calculated as described in Section \ref{subsec:weightedrate}. Expressed in per cent.
        \item (7) --- The unweighted binomial probability (per cent) of making zero detections in our sample of 100 dwarf galaxies, assuming an intrinsic rate of $R_0$.
        \item (8) --- The weighted binomial probability (per cent) of making zero detections in our sample of 100 dwarf galaxies, assuming an intrinsic rate of $R_w$.
    \end{tablenotes}
\end{table*}

To further illustrate this, we plot the distribution of dwarf galaxies and the MCP--NSA detection rate for intrinsic and corrected [O {\footnotesize III}] flux in the top and bottom panels of Figure \ref{fig:O3flux}, respectively. The trend in [O {\footnotesize III}] flux is largely the same as in luminosity: Our sample is both less obscured and [O {\footnotesize III}]-faint relative to previous surveys. The [O {\footnotesize III}] flux-weighted detection rate is $R_w=1.57^{+0.56}_{-0.42}\%$ for observed flux and $R_w=0.65^{+0.27}_{-0.21}\%$ for extinction-corrected flux. The latter is much lower than $R_w$ weighted by any other property, and corresponds to $P_w(0)=52.15^{+11.01}_{-14.34}\%$. We checked whether this was a consequence of removing galaxies with poorly-constrained distances. Since flux does not depend on distance, we can re-calculate $R_w$ and $P_w(0)$ using all galaxies in the MCP--NSA sample regardless of ZDIST quality. We do so and find that $R_w$ and $P_w(0)$ agree within errors with the sample where galaxies with unreliable ZDIST are removed.

\section{Conclusions} \label{sec:conclusion}

We used the GBT to search for water maser emission in 100 dwarf galaxies with optical spectroscopic signatures of accretion onto a massive BH, the largest such survey to date. We detected no new masers down to a 5$\sigma$ limit of $\sim$12 mJy. Assuming Poisson statistics, the 95\% Confidence Level upper limit of our zero counts among 100 targets is $\sim$3, meaning we cannot as yet rule out an intrinsic maser rate of 3\%, as seen in more massive galaxies. The 95\% CL upper limit for zero counts in 127 galaxies, including those with archival GBT spectra, is 2.4\%, but we note that the average noise in the archival spectra was $\sim$50\% higher than in our new observations.

We compared our target dwarf galaxies with the detection rate in archival observations of galaxies in the NSA, weighted by the distributions of BPT classification, Balmer decrement, total galaxy stellar mass, [O {\footnotesize III}] $\lambda5007$ flux, and [O {\footnotesize III}] $\lambda5007$ luminosity of galaxies in our target sample. We found that the dwarf sample is comprised of significantly fewer Seyfert 2 galaxies, is less obscured, and is [O {\footnotesize III}]-faint relative to previous surveys for maser emission. The binomial probability of making zero detections in 100 galaxies is as high as $\sim$20--50\% when detection rate is weighted by the host galaxy optical properties, compared with a probability of $\sim$5--10\% using the unweighted rate.

We also see an increase in detection rate with increasing $M_*$, but this trend is less pronounced than the trend in the other properties we studied. Given that the targets in our survey span a range of AGN properties that are associated with a lower detection rate even in more massive galaxies, further observations are required to discern whether there is an intrinsic difference between the maser fraction in active dwarf galaxies and in their more massive counterparts with the same optical properties. Despite the lower detection rate, the existence of the anomalous $\sim$10$^5~M_{\sun}$ BH in the low-mass galaxy IC 750 motivates a continued search for these rare objects.

\section*{Acknowledgements}

We thank the anonymous referee for their helpful comments. We thank Joseph Gelfand and Lincoln J. Greenhill for their helpful discussions.

This work was made possible by the use of the Robert C. Byrd Green Bank Telescope at the Green Bank Observatory. The Green Bank Observatory is a facility of the National Science Foundation operated under cooperative agreement by Associated Universities, Inc.

Funding for SDSS-III has been provided by the Alfred P. Sloan Foundation, the Participating Institutions, the National Science Foundation, and the U.S. Department of Energy Office of Science. The SDSS-III web site is http://www.sdss3.org/.

SDSS-III is managed by the Astrophysical Research Consortium for the Participating Institutions of the SDSS-III Collaboration including the University of Arizona, the Brazilian Participation Group, Brookhaven National Laboratory, Carnegie Mellon University, University of Florida, the French Participation Group, the German Participation Group, Harvard University, the Instituto de Astrofisica de Canarias, the Michigan State/Notre Dame/JINA Participation Group, Johns Hopkins University, Lawrence Berkeley National Laboratory, Max Planck Institute for Astrophysics, Max Planck Institute for Extraterrestrial Physics, New Mexico State University, New York University, Ohio State University, Pennsylvania State University, University of Portsmouth, Princeton University, the Spanish Participation Group, University of Tokyo, University of Utah, Vanderbilt University, University of Virginia, University of Washington, and Yale University.

This research made use of ASTROPY, a community-developed core PYTHON package for Astronomy \citep{Astropy};  IPYTHON \citep{iPython}; MATPLOTLIB \citep{matplotlib}; NUMPY \citep{numpy}; and TOPCAT \citep{TOPCAT}.

\section*{Data Availability Statement}

The new data underlying this article are available in the article. We also use publicly available data from the NSA (\url{http://nsatlas.org/}), the MCP (\url{https://safe.nrao.edu/wiki/bin/view/Main/MegamaserCosmologyProject}) and \citet{Reines13}.

%%%%%%%%%%%%%%%%%%%%%%%%%%%%%%%%%%%%%%%%%%%%%%%%%%



\bibliographystyle{mnras}

%%%%%%%%%%%%%%%%%%%%%%%%%%%%%%%%%%%%%%%%%%%%%%%%%%

%%%%%%%%%%%%%%%%% APPENDICES %%%%%%%%%%%%%%%%%%%%%

\appendix

\section{Non-detections in our survey} \label{sec:nondetections}

\begin{figure*}
	\includegraphics[width=0.98\linewidth]{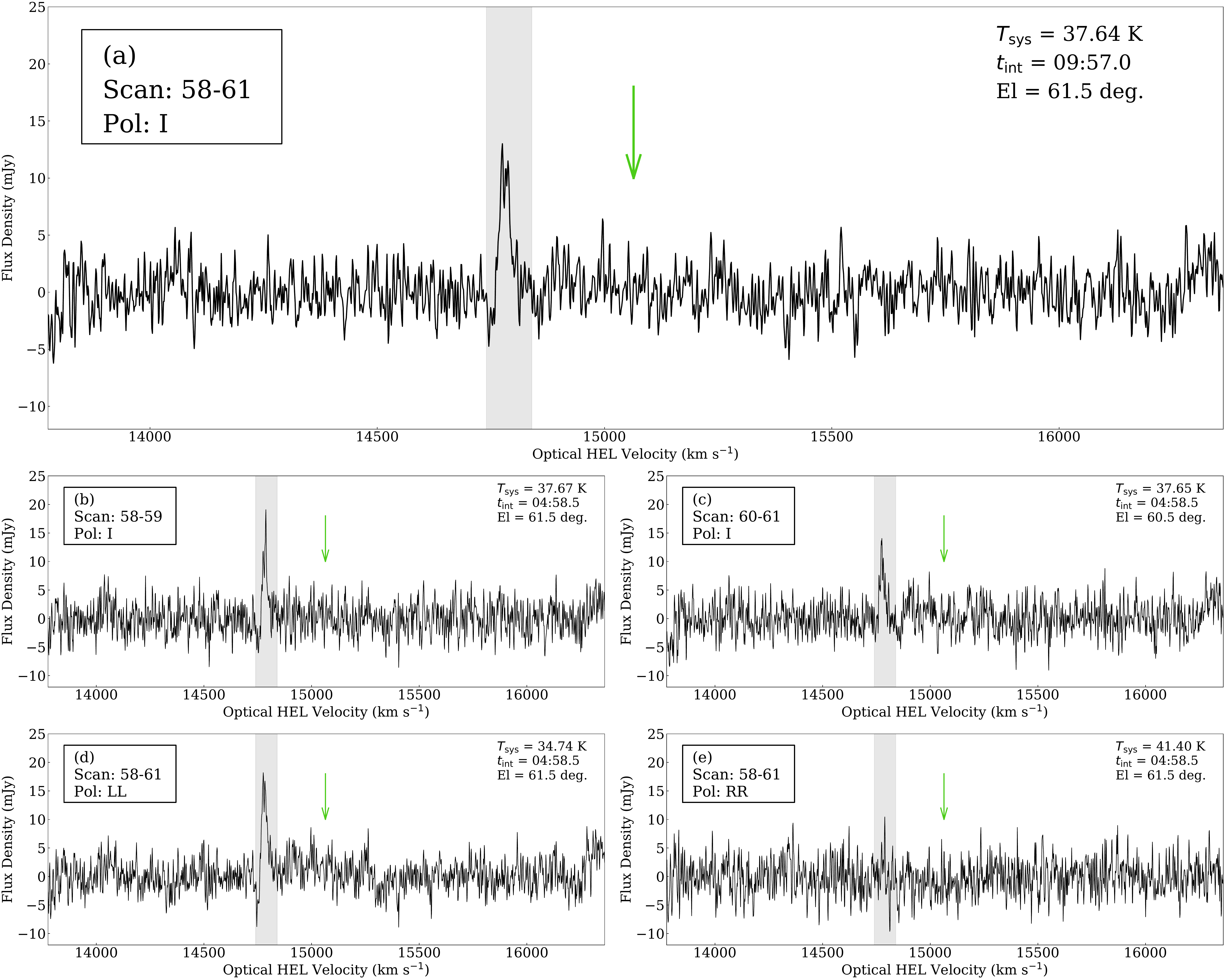}
	\caption{GBT spectra of the galaxy SDSS J114359.58+244251.7. All panels are for observations from 2019-Nov-01, for session AGBT19B\_281\_01. The scans and polarizations for the displayed spectra are given in the top left corner of each panel. The average system temperatures, total integration times, and elevations at the start of observation for the displayed spectra are shown in the top right corner of each panel. Panel (a) shows the averaged spectrum for both nods and both polarizations. Panels (b) and (c) are for the first and second individual nods, respectively, still with both polarizations averaged. Panels (d) and (e) show the spectra for both nods in the left-handed and right-handed polarization, respectively. In all panels, the SDSS-measured systemic velocity of the galaxy of 15064 km s$^{-1}$ is marked by the green arrow, and the shaded region from 14740--14840 km s$^{-1}$ highlights the location of a false detection of broad emission. The bottom two panels clearly show that this feature is present only in one polarization, and therefore not real maser emission. }  \label{fig:spectra}
\end{figure*}

In this appendix we discuss the non-detections in our survey. We devote particular attention to an initial false positive in the spectrum of SDSS J114359.58+244251.7 (RGG 18), in scans 58--61 of session AGBT19B\_281\_01. This spectrum initially appeared to have a broad ($\sim$25 km s$^{-1}$) feature, blueshifted $\sim$300 km s$^{-1}$ from the systemic velocity of the host galaxy in its observed spectrum. However, upon inspection of each polarization separately, we found this feature was only detected in the left-handed polarization, with a maximum flux density of 18.2 mJy (after Hanning smoothing), while the right-handed polarization had a maximum flux density of 10.5 mJy, but no discernible emission features, corresponding to a difference of $\approx$3--4$\sigma$. We show the spectrum for this galaxy resulting from the data reduction procedure in Section \ref{sec:data} in the top panel of Figure \ref{fig:spectra}. The remaining panels of Figure \ref{fig:spectra} are also for SDSS J114359.58+244251.7, but separately show the spectra for each individual nod, and again for each individual polarization. We checked the spectrum of the known maser NGC 4258, which we observed for a single nod of 60 seconds, for a similar disparity between its polarizations. NGC 4258 is an extremely strong source, with a maximum 22 GHz flux density of $\sim$8--15 Jy between November 2019 and February 2020, so a one-minute nod was sufficient to measure its peak flux density. The source had a maximum of 10.7 Jy in its left-handed spectrum and 12.0 Jy in its right-handed spectrum, indicating that there is no real maser emission from SDSS J114359.58+244251.7. A follow-up 1-hour integration on the same source (project code AGBT20A\_419) found no emission, so we assuredly classify it as a non-detection.

In Table \ref{tab:nondetections}, we give the details of our observations of each target, including that of SDSS J114359.58+244251.7, as well as of the re-reduced archival GBT spectra.

\begin{table*} 
\caption{Summary of non-detections} \label{tab:nondetections}
\begin{tabular}{rccccrccccccc}
\hline
\hline
(1) & (2) & (3) & (4) & (5) & (6) & (7) & (8) & (9) & (10) & (11) & (12) & (13) \\
RGG & Name & BPT & RA & Dec & $v_{\rm{sys}}$ & Date & $t_{\rm{int}}$ & El. & $T_{\rm{sys}}$ & $\sigma_{\rm{BC}}$ & $\sigma_{\rm{BC+Han}}$ & $\Delta v$ \\
& & & & & km s$^{-1}$ & & s & deg. & K & mJy & mJy & km s$^{-1}$ \\
\hline
\multicolumn{13}{c}{This Work} \\ 
\hline 
--- &     Was 49b & Sy (--) & 12:14:17.80 &  +29:31:43.0 & 19187 & 2019-11-01 &  597.2 & 60.2 & 38.56 & 3.2 & 2.1 & 1.4 \\
 13 &  J1022+4642 &   L (L) & 10:22:52.21 &  +46:42:20.7 & 11752 & 2019-12-07 &  594.4 & 65.8 & 38.36 & 3.1 & 1.9 & 1.3 \\
 17 &  J1143+2608 & Sy (Sy) & 11:43:02.42 &  +26:08:19.0 &  6927 & 2019-11-01 &  597.0 & 64.2 & 40.90 & 3.4 & 2.1 & 1.3 \\
 22 &  J1304+0755 & Sy (Sy) & 13:04:34.92 &  +07:55:05.0 & 14045 & 2020-02-08 &  592.1 & 59.3 & 43.76 & 3.6 & 2.4 & 1.4 \\
 26 &  J1349+4202 & Sy (Sy) & 13:49:39.37 &  +42:02:41.4 & 12330 & 2019-12-07 &  594.5 & 79.4 & 37.63 & 3.1 & 1.9 & 1.3 \\
 28 &  J1405+1146 & Sy (Sy) & 14:05:10.39 &  +11:46:16.9 &  5171 & 2019-12-11 &  593.5 & 62.7 & 39.37 & 3.3 & 2.1 & 1.3 \\
 29 &  J1412+1029 & Sy (Sy) & 14:12:08.47 &  +10:29:53.8 &  9613 & 2019-12-11 &  593.6 & 61.2 & 39.73 & 3.2 & 2.0 & 1.3 \\
 30 &  J1420+2242 & Sy (Sy) & 14:20:44.94 &  +22:42:36.8 &  9062 & 2020-02-08 &  593.1 & 58.5 & 46.47 & 3.9 & 2.4 & 1.3 \\
 33 &  J1447+1339 &   L (L) & 14:47:12.80 &  +13:39:39.2 &  9526 & 2020-02-08 &  592.6 & 59.1 & 45.73 & 3.7 & 2.3 & 1.3 \\
 34 &  J1539+1714 & Sy (Sy) & 15:39:41.65 &  +17:14:22.0 & 13416 & 2020-02-09 &  593.5 & 55.0 & 46.29 & 3.9 & 2.5 & 1.3 \\
 35 &  J1540+3155 & Sy (Sy) & 15:40:59.61 &  +31:55:07.3 & 15799 & 2019-11-02 &  596.6 & 76.7 & 41.04 & 3.5 & 2.2 & 1.4 \\
 37 & J0100--0110 &   C (C) & 01:00:05.93 & --01:10:59.0 & 15423 & 2019-12-08 &  592.1 & 50.2 & 44.59 & 4.2 & 3.0 & 1.4 \\
 38 &  J0246+0007 &   A (C) & 02:46:35.04 &  +00:07:18.7 &  8582 & 2019-12-08 &  592.2 & 43.8 & 49.74 & 4.1 & 2.5 & 1.3 \\
 39 & J0306--0024 &   C (C) & 03:06:44.61 & --00:24:31.7 &  7544 & 2019-12-08 &  592.3 & 42.2 & 51.01 & 4.5 & 2.8 & 1.3 \\
 40 &  J0748+5100 &   C (C) & 07:48:29.23 &  +51:00:52.2 &  5705 & 2019-12-07 &  595.2 & 48.9 & 42.17 & 3.7 & 2.3 & 1.3 \\
 41 &  J0802+2030 &   C (C) & 08:02:28.84 &  +20:30:50.3 &  8588 & 2019-12-08 &  592.9 & 57.5 & 48.76 & 4.3 & 2.8 & 1.3 \\
 42 &  J0810+0733 &   C (C) & 08:10:10.69 &  +07:33:37.2 & 15670 & 2019-12-12 &  592.5 & 58.8 & 41.87 & 3.8 & 2.7 & 1.4 \\
 43 &  J0813+3108 &   C (C) & 08:13:53.46 &  +31:08:24.4 & 14396 & 2019-12-08 &  593.6 & 61.0 & 45.03 & 4.0 & 2.7 & 1.4 \\
 44 &  J0815+2547 &   C (C) & 08:15:49.09 &  +25:47:01.4 &  7435 & 2019-12-08 &  593.1 & 61.0 & 48.30 & 4.1 & 2.5 & 1.3 \\
 45 &  J0820+3025 &   C (C) & 08:20:13.92 &  +30:25:03.0 &  5911 & 2019-12-08 &  593.5 & 59.7 & 47.40 & 4.0 & 2.5 & 1.3 \\
 46 &  J0837+6013 &   C (C) & 08:37:40.88 &  +60:13:39.3 &  8072 & 2019-12-07 &  595.5 & 53.4 & 41.60 & 3.4 & 2.1 & 1.3 \\
 47 &  J0844+2818 &   C (C) & 08:44:49.13 &  +28:18:53.5 &  6082 & 2019-12-08 &  593.5 & 60.3 & 48.19 & 4.0 & 2.4 & 1.3 \\
 49 &  J0855+3545 &   A (C) & 08:55:48.25 &  +35:45:02.0 & 15383 & 2019-12-08 &  594.1 & 60.8 & 45.04 & 3.9 & 2.6 & 1.4 \\
 51 &  J0907+3310 &   C (C) & 09:07:46.23 &  +33:10:00.8 & 14415 & 2019-12-08 &  593.8 & 64.6 & 45.06 & 4.2 & 3.1 & 1.4 \\
 52 &  J0915+1006 &   C (C) & 09:15:55.31 &  +10:06:07.5 &  9120 & 2019-12-12 &  592.5 & 57.3 & 44.50 & 3.8 & 2.5 & 1.3 \\
 54 &  J0927+3312 &   C (C) & 09:27:47.65 &  +33:12:33.1 & 11031 & 2019-12-12 &  593.5 & 65.0 & 43.74 & 3.6 & 2.3 & 1.3 \\
 55 &  J0932+4205 &   C (C) & 09:32:26.79 &  +42:05:16.8 &  7664 & 2019-12-07 &  594.3 & 59.8 & 40.56 & 3.4 & 2.1 & 1.3 \\
 56 &  J0932+5115 &   C (C) & 09:32:39.45 &  +51:15:42.9 & 13845 & 2019-12-07 &  594.7 & 61.5 & 39.09 & 3.2 & 2.0 & 1.4 \\
 57 &  J0934+2640 &   C (C) & 09:34:44.94 &  +26:40:28.2 &  9526 & 2019-12-12 &  593.1 & 65.5 & 44.65 & 3.7 & 2.3 & 1.3 \\
 58 &  J0938+0631 &  Sy (C) & 09:38:21.54 &  +06:31:30.8 &  6640 & 2019-12-12 &  592.6 & 52.9 & 45.92 & 3.8 & 2.4 & 1.3 \\
 59 &  J0947+0501 &   C (C) & 09:47:05.72 &  +05:01:59.8 &  7167 & 2019-12-12 &  592.7 & 51.8 & 46.35 & 3.9 & 2.4 & 1.3 \\
 60 &  J0950+3604 &   C (C) & 09:50:20.07 &  +36:04:46.6 &  6464 & 2019-12-12 &  593.7 & 66.9 & 45.20 & 3.8 & 2.4 & 1.3 \\
 61 &  J0951+0307 &  SF (C) & 09:51:47.61 &  +03:07:22.0 &  6141 & 2019-12-12 &  592.6 & 51.3 & 46.26 & 3.8 & 2.3 & 1.3 \\
 63 &  J1000+0059 &   C (C) & 10:00:08.40 &  +00:59:05.2 &  9381 & 2019-12-12 &  592.7 & 49.6 & 46.12 & 3.8 & 2.3 & 1.3 \\
 64 &  J1004+2313 &   C (C) & 10:04:23.33 &  +23:13:23.4 &  7868 & 2019-12-12 &  592.8 & 70.0 & 44.84 & 3.7 & 2.2 & 1.3 \\
 65 &  J1011+2334 &   C (C) & 10:11:03.34 &  +23:34:47.2 &  6610 & 2019-12-12 &  592.7 & 72.5 & 44.56 & 3.7 & 2.3 & 1.3 \\
 66 &  J1017+3932 &   C (C) & 10:17:47.09 &  +39:32:07.7 & 15752 & 2019-12-07 &  594.2 & 62.6 & 38.26 & 3.3 & 2.2 & 1.4 \\
 68 &  J1023+0654 &   C (C) & 10:23:55.74 &  +06:54:52.7 &  9874 & 2019-12-12 &  592.9 & 53.9 & 45.12 & 3.8 & 2.4 & 1.3 \\
 69 &  J1028+1845 &   C (C) & 10:28:33.33 &  +18:45:13.9 &  8101 & 2019-12-12 &  592.7 & 70.2 & 44.89 & 3.8 & 2.3 & 1.3 \\
 70 &  J1038+3802 &   C (C) & 10:38:48.93 &  +38:08:45.2 & 10713 & 2020-02-09 & 1190.8 & 41.8 & 51.02 & 3.2 & 2.1 & 1.3 \\
 71 &  J1041+1436 &   C (C) & 10:41:19.80 &  +14:36:41.1 & 14444 & 2019-12-12 &  592.8 & 65.8 & 43.24 & 3.7 & 2.4 & 1.4 \\
 72 &  J1046+2555 &   C (C) & 10:46:44.86 &  +25:55:02.0 &  6024 & 2019-12-12 &  593.1 & 70.2 & 45.09 & 3.9 & 2.5 & 1.3 \\
 73 &  J1047+3355 &   A (C) & 10:47:58.09 &  +33:55:36.8 &  6757 & 2019-12-07 &  594.0 & 62.5 & 39.82 & 3.3 & 2.0 & 1.3 \\
 74 &  J1050+1630 &   C (C) & 10:50:06.07 &  +16:30:53.2 & 15298 & 2019-12-12 &  593.0 & 66.5 & 42.57 & 3.7 & 2.5 & 1.4 \\
\hline
\end{tabular}
\tablenotes{
\item Galaxies are ordered by \citet{Reines13} ID number. We list first all of our own observations, and then non-detections from the archival data.
\item (1) --- ID from \citet{Reines13}, except for Was 49b, which is not from this sample.
\item (2) --- Abbreviated IAU name.
\item (3) --- BPT classification, using both [N II]/H$\alpha$ and [S II]/H$\alpha$ diagnostics as described in Section \ref{subsec:BPT}. Possible values are: Sy--Seyfert, L--LINER, C--Composite, A--Ambiguous, SF--Star-forming. The value in parentheses gives the classification according to emission line fluxes from \citet{Reines13}, rather than from the NSA.
\item (4) --- Right ascension (J2000).
\item (5) --- Declination (J2000).
\item (6) --- Heliocentric systemic velocity. For observations from this work, we used the preferred value from NED at the time of observation.
\item (7) --- Observation date (year-month-day).
\item (8) --- Total on-source integration time, in seconds.
\item (9) --- Source elevation at the start of the first nod, in degrees.
\item (10) --- Average system temperature in Kelvin over the duration of the observation of a given source.
\item (11) --- Line-free RMS noise after boxcar smoothing.
\item (12) --- Line-free RMS noise after boxcar smoothing \textit{and} Hanning smoothing.
\item (13) --- Average channel width in km s$^{-1}$ after boxcar smoothing.
}
\end{table*}

\begin{table*}
\contcaption{Summary of non-detections}
\begin{tabular}{rccccrccccccc}
\hline
\hline
RGG & Name & BPT & RA & Dec & $v_{\rm{sys}}$ & Date & $t_{\rm{int}}$ & El. & $T_{\rm{sys}}$ & $\sigma_{\rm{BC}}$ & $\sigma_{\rm{BC+Han}}$ & $\Delta v$ \\
& & & & & km s$^{-1}$ & & s & deg. & K & mJy & mJy & km s$^{-1}$ \\
\hline 
\multicolumn{13}{c}{This Work (\textit{continued})} \\ 
\hline 
 75 &  J1055+1356     &  C (C) & 10:55:55.12 &  +13:56:16.9 &  7050 & 2019-12-12 & 592.9 & 62.6     & 45.45     & 3.7 & 2.3 & 1.3 \\
 76 &  J1100+4957     &  C (C) & 11:00:27.04 &  +49:57:45.1 &  7478 & 2019-11-02 & 597.1 & 75.6     & 44.58     & 3.7 & 2.3 & 1.3 \\
 77 &  J1102+1231     &  C (C) & 11:02:19.75 &  +12:31:07.1 &  7985 & 2019-12-12 & 593.0 & 59.4     & 45.87     & 3.8 & 2.4 & 1.3 \\
 78 &  J1126+0955     & SF (C) & 11:26:31.77 &  +09:55:15.4 &  9236 & 2019-12-12 & 593.5 & 51.6     & 46.33     & 4.0 & 2.5 & 1.3 \\
 79 &  J1129+6538     & Sy (C) & 11:29:57.63 &  +65:38:04.7 & 13165 & 2019-11-02 & 597.5 & 62.2     & 41.93     & 3.4 & 2.1 & 1.3 \\
 80 &  J1130+2322     &  C (C) & 11:30:24.27 &  +23:22:00.1 &  7184 & 2019-11-01 & 597.1 & 56.1     & 42.52     & 3.5 & 2.1 & 1.3 \\
 81 &  J1131+3509     &  C (C) & 11:31:29.20 &  +35:09:58.9 &  9932 & 2019-12-07 & 593.9 & 69.2     & 39.05     & 3.2 & 2.0 & 1.3 \\
 82 &  J1134+0018     &  C (C) & 11:34:13.74 &  +00:18:34.0 &  8305 & 2019-12-12 & 593.8 & 40.7     & 49.69     & 4.3 & 2.6 & 1.3 \\
 83 &  J1148+2749     &  C (C) & 11:44:00.66 &  +27:49:46.2 &  8824 & 2019-11-01 & 596.9 & 68.1     & 40.22     & 3.3 & 2.0 & 1.3 \\
 84 &  J1149+1851     &  C (C) & 11:49:02.63 &  +18:51:11.5 &  5700 & 2019-12-07 & 593.5 & 60.2     & 39.55     & 3.3 & 2.0 & 1.3 \\
 85 &  J1153+3011     &  C (C) & 11:53:08.65 &  +30:11:14.8 &  6441 & 2019-11-01 & 597.1 & 60.7     & 42.65     & 3.5 & 2.2 & 1.3 \\
 86 &  J1153+1308     &  C (C) & 11:53:59.06 &  +13:08:53.6 &  6698 & 2019-12-12 & 592.8 & 64.2     & 46.00     & 3.8 & 2.3 & 1.3 \\
 87 &  J1157+2807     &  C (C) & 11:57:17.47 &  +28:07:08.5 &  6551 & 2019-11-01 & 597.1 & 58.4     & 43.47     & 3.6 & 2.2 & 1.3 \\
 88 &  J1158+5753     &  C (C) & 11:58:12.53 &  +57:53:22.1 & 12438 & 2019-11-02 & 597.3 & 70.6     & 41.73     & 3.5 & 2.1 & 1.3 \\
 90 &  J1211+4658     &  C (C) & 12:11:55.71 &  +46:58:54.5 &   967 & 2019-11-02 & 597.0 & 79.4     & 44.03     & 3.7 & 2.2 & 1.2 \\
 91 &  J1218+2004     &  C (C) & 12:18:13.43 &  +20:04:36.8 & 13387 & 2019-12-07 & 593.6 & 63.9     & 37.59     & 3.1 & 1.9 & 1.3 \\
 93 &  J1220+2920     &  C (C) & 12:20:42.09 &  +29:20:45.2 &   644 & 2019-11-01 & 597.2 & 59.2     & 47.72     & 4.0 & 2.4 & 1.2 \\
 95 &  J1228+0926     &  C (C) & 12:28:15.93 &  +09:26:10.7 &   359 & 2019-12-12 & 592.8 & 60.9     & 46.48     & 3.8 & 2.3 & 1.2 \\
 96 &  J1240+0927     &  C (C) & 12:40:33.86 &  +09:27:28.4 &  8218 & 2019-12-12 & 593.0 & 60.9     & 46.56     & 3.9 & 2.5 & 1.3 \\
 97 &  J1249+0518     &  C (C) & 12:49:57.86 &  +05:18:41.0 &   658 & 2019-12-12 & 592.9 & 56.8     & 47.46     & 4.0 & 2.5 & 1.2 \\
 98 &  J1309+4154     &  C (C) & 13:09:16.09 &  +41:54:49.4 &  8787 & 2019-12-07 & 594.6 & 74.2     & 39.89     & 3.4 & 2.1 & 1.3 \\
 99 &  J1310+3000     &  C (C) & 13:10:07.10 &  +30:00:56.1 & 10611 & 2019-11-01 & 597.0 & 66.7     & 42.39     & 3.5 & 2.2 & 1.3 \\
100 &  J1321+3844     &  C (C) & 13:21:56.40 &  +38:44:04.9 &   975 & 2019-12-07 & 594.2 & 78.9     & 39.22     & 3.2 & 1.9 & 1.2 \\
101 &  J1325+3153     & SF (C) & 13:25:32.34 &  +31:53:33.1 & 11146 & 2019-12-07 & 593.9 & 80.9     & 37.81     & 3.0 & 1.9 & 1.3 \\
103 &  J1341+2428     &  C (C) & 13:41:30.13 &  +24:28:52.0 &  8393 & 2019-12-11 & 593.7 & 74.0     & 39.12     & 3.3 & 2.0 & 1.3 \\
104 &  J1341+5006     &  C (C) & 13:41:56.15 &  +50:06:42.2 & 11284 & 2019-11-02 & 597.0 & 76.1     & 42.16     & 3.3 & 2.0 & 1.3 \\
105 &  J1351+4012$^a$ &  C (C) & 13:51:25.35 &  +40:12:47.8 &  2466 &    ---$^a$ & 594.3 & 83.2$^a$ & 38.99$^a$ & 3.3 & 2.1 & 1.3 \\
106 &  J1401+5152     &  L (C) & 14:01:16.05 &  +51:52:22.8 &  2270 & 2019-11-02 & 597.0 & 75.0     & 44.26     & 3.7 & 2.3 & 1.3 \\
107 &  J1407+4705     &  C (C) & 14:07:27.52 &  +47:05:23.8 & 12897 & 2019-11-02 & 597.1 & 73.0     & 41.03     & 3.4 & 2.1 & 1.3 \\
108 &  J1407+5032     &  C (C) & 14:07:35.48 &  +50:32:42.8 &  2048 & 2019-11-02 & 597.0 & 74.3     & 43.95     & 3.6 & 2.2 & 1.3 \\
109 & J1411--0029     &  C (C) & 14:11:20.29 & --00:29:50.7 &  7459 & 2020-02-08 & 592.1 & 48.2     & 49.12     & 4.1 & 2.5 & 1.3 \\
110 &  J1414+2004     &  C (C) & 14:14:38.18 &  +20:04:05.8 &  8480 & 2020-02-08 & 592.6 & 61.9     & 47.08     & 4.0 & 2.5 & 1.3 \\
111 &  J1427+1100$^b$ & SF (C) & 14:27:20.32 &  +11:00:53.0 &  8130 &    ---$^b$ & 593.0 & 60.2$^b$ & 43.13$^b$ & 3.6 & 2.2 & 1.3 \\
112 &  J1431+2616     & Sy (C) & 14:31:46.75 &  +26:16:24.3 &  4610 & 2020-02-08 & 593.5 & 56.3     & 48.29     & 4.2 & 2.7 & 1.3 \\
113 &  J1438+5107     &  C (C) & 14:38:59.15 &  +51:07:14.5 &  2223 & 2019-11-02 & 596.9 & 73.7     & 44.41     & 3.8 & 2.3 & 1.3 \\
114 &  J1442+2054     & Sy (C) & 14:42:52.78 &  +20:54:51.6 & 12528 & 2020-02-08 & 593.3 & 55.5     & 44.79     & 3.8 & 2.4 & 1.3 \\
115 &  J1455+0816     &  C (C) & 14:55:18.50 &  +08:16:06.4 & 10627 & 2020-02-08 & 592.5 & 54.9     & 46.76     & 3.8 & 2.4 & 1.3 \\
116 &  J1514+0603     &  C (C) & 15:14:22.89 &  +06:03:08.3 &  7050 & 2020-02-08 & 592.7 & 52.6     & 49.63     & 4.2 & 2.6 & 1.3 \\
117 &  J1516+0845     &  C (C) & 15:16:08.80 &  +08:45:31.9 & 10309 & 2020-02-08 & 592.7 & 56.1     & 47.78     & 4.0 & 2.5 & 1.3 \\
118 &  J1523+1145     &  C (C) & 15:23:03.80 &  +11:45:46.0 &  7196 & 2020-02-08 & 592.7 & 60.4     & 46.76     & 3.9 & 2.4 & 1.3 \\
120 &  J1529+0830     &  C (C) & 15:29:13.46 &  +08:30:10.6 & 12614 & 2020-02-08 & 592.8 & 55.7     & 45.03     & 3.7 & 2.3 & 1.3 \\
121 &  J1530+1246     &  C (C) & 15:30:31.44 &  +12:46:54.3 & 15440 & 2020-02-08 & 592.9 & 61.7     & 43.26     & 3.6 & 2.2 & 1.4 \\
122 &  J1530+3136     &  C (C) & 15:30:56.86 &  +31:36:49.5 & 10954 & 2019-11-02 & 596.5 & 76.3     & 42.77     & 3.5 & 2.2 & 1.3 \\
124 &  J1536+1626     &  C (C) & 15:36:31.00 &  +16:26:25.6 &  1882 & 2020-02-09 & 593.2 & 56.0     & 49.15     & 4.1 & 2.5 & 1.2 \\
125 &  J1547+3413     &  C (C) & 15:47:03.21 &  +34:13:19.4 & 11850 & 2019-11-02 & 596.7 & 71.2     & 42.53     & 3.3 & 2.0 & 1.3 \\
126 &  J1602+1429     &  C (C) & 16:02:30.77 &  +14:29:12.9 &  9903 & 2020-02-09 & 593.7 & 51.2     & 51.35     & 4.2 & 2.6 & 1.3 \\
128 &  J1605+0850     & Sy (C) & 16:05:44.58 &  +08:50:43.9 &  4581 & 2020-02-09 & 593.7 & 47.6     & 51.01     & 4.2 & 2.6 & 1.3 \\
129 &  J1609+5350     & SF (C) & 16:09:58.68 &  +53:50:11.3 &  5268 & 2019-12-11 & 595.9 & 59.1     & 38.91     & 3.3 & 2.0 & 1.3 \\
130 &  J1621+4023     &  C (C) & 16:21:04.15 &  +40:23:22.4 &  9999 & 2019-11-02 & 596.9 & 63.6     & 44.35     & 3.6 & 2.2 & 1.3 \\
131 &  J1625+4843     &  C (C) & 16:25:14.66 &  +48:43:16.8 &  6233 & 2019-11-02 & 597.2 & 60.8     & 45.04     & 3.7 & 2.3 & 1.3 \\
132 &  J1632+2901     &  C (C) & 16:32:10.87 &  +29:01:30.6 & 11090 & 2019-11-02 & 596.6 & 68.0     & 42.06     & 3.6 & 2.1 & 1.3 \\
133 &  J1701+2552     &  C (C) & 17:01:04.47 &  +25:52:08.0 & 11579 & 2020-02-09 & 594.6 & 55.6     & 50.90     & 4.2 & 2.6 & 1.3 \\
134 &  J1718+6035     &  C (C) & 17:18:30.90 &  +60:35:07.0 &  3871 & 2019-12-11 & 596.8 & 50.3     & 40.66     & 3.4 & 2.1 & 1.3 \\
135 &  J1732+5958     &  C (C) & 17:32:02.96 &  +59:58:55.0 &  8597 & 2019-12-11 & 596.8 & 50.1     & 40.70     & 3.4 & 2.0 & 1.3 \\
136 & J2356--0024     &  C (C) & 23:56:09.18 & --00:24:28.7 &  7674 & 2019-12-08 & 592.1 & 50.7     & 47.07     & 4.0 & 2.4 & 1.3 \\
\hline
\end{tabular}
\tablenotes{
\item \textit{a} --- J1351+4012 was observed for 5 minutes of integration on 2019-12-07 and 5 minutes of integration on 2019-12-11. We list the mean $T_{\rm{sys}}$ and elevation for these scans. RMS values are from the averaged, 10-minute spectra.  
\item \textit{b} --- J1427+1000 was observed for 5 minutes of integration on 2019-12-11 and 5 minutes of integration on 2020-02-08. We list the mean $T_{\rm{sys}}$ and elevation for these scans. RMS values are from the averaged, 10-minute spectra. 
}
\end{table*}

\begin{table*}
\contcaption{Summary of non-detections}
\begin{tabular}{rccccrccccccc}
\hline
\hline
RGG & Name & BPT & RA & Dec & $v_{\rm{sys}}$ & Date & $t_{\rm{int}}$ & El. & $T_{\rm{sys}}$ & $\sigma_{\rm{BC}}$ & $\sigma_{\rm{BC+Han}}$ & $\Delta v$ \\
& & & & & km s$^{-1}$ & & s & deg. & K & mJy & mJy & km s$^{-1}$ \\
\hline 
\multicolumn{13}{c}{Archival Data} \\ 
\hline
  2 & J0248--0025 & Sy (Sy) & 02:48:25.25 & --00:25:41.4 &  7399 & 2014-11-04 & 584.4     & 47.6 & 74.95     &     7.4 &     4.5 &     1.4 \\
  3 &  J0322+4011 & Sy (Sy) & 03:22:24.64 &  +40:11:19.9 &  7825 & 2014-12-15 & 584.6     & 74.6 & 47.42     &     4.6 &     2.9 &     1.4 \\
  4 &  J0811+2328 & Sy (Sy) & 08:11:45.31 &  +23:28:25.7 &  4720 & 2014-11-04 & 584.7     & 64.1 & 67.21     &     6.8 &     4.2 &     1.4 \\
  5 &  J0823+0313 &  A (Sy) & 08:23:34.84 &  +03:13:15.7 &  2931 & 2014-11-04 & 584.5     & 54.6 & 70.62     &     7.3 &     4.3 &     1.3 \\
  6 &  J0840+1818 & Sy (Sy) & 08:48:25.54 &  +18:18:58.9 &  4495 & 2014-11-04 & 584.7     & 62.3 & 64.17     &     6.4 &     4.0 &     1.4 \\
  7 &  J0842+4039 & Sy (Sy) & 08:42:04.91 &  +40:39.34.4 &  8825 & 2014-11-13 & 584.8     & 81.8 & 59.98     &     6.3 &     4.2 &     1.4 \\
  8 &  J0902+1410 &   L (L) & 09:02:22.77 &  +14:10:49.2 &  8824 & 2014-11-05 & 584.6     & 62.0 & 70.23     &     7.0 &     4.3 &     1.4 \\
 10 &  J0921+2131 & Sy (Sy) & 09:21:30.00 &  +21:31:39.4 &  9377 & 2015-01-23 & 584.6     & 55.9 & 39.84     &     4.0 &     2.6 &     1.4 \\
 12 &  J1009+2656 & Sy (Sy) & 10:09:35.69 &  +26:56:48.9 &  4306 & 2014-11-05 & 584.7     & 75.2 & 80.29     &     7.8 &     5.0 &     1.4 \\
 14 &  J1105+2241 & Sy (Sy) & 11:05:03.99 &  +22:41:23.6 &  7433 & 2015-01-09 & 584.5     & 74.1 & 53.47     &     5.4 &     3.3 &     1.4 \\
 15 &  J1109+6123 & Sy (Sy) & 11:09:12.38 &  +61:23:47.1 &  2025 & 2015-03-12 & 150.0$^c$ & 65.0 & 39.70     & ---$^c$ & 6.3$^c$ &     1.3 \\
 16 &  J1113+0444 & Sy (Sy) & 11:13:19.24 &  +04:44:25.4 &  7937 & 2015-01-08 & 584.4     & 55.2 & 34.15     &     3.3 &     2.0 &     1.4 \\
 19 &  J1144+3340 & Sy (Sy) & 11:44:18.86 &  +33:40:07.6 &  9762 & 2015-02-16 & 593.0     & 74.1 & 36.82     &     3.2 &     2.1 &     1.3 \\
 23 &  J1304+3626 & Sy (Sy) & 13:04:57.85 &  +36:26:22.3 &  6862 & 2015-04-22 & 595.6     & 38.6 & 79.09     &     7.2 &     4.8 &     1.3 \\
 24 &  J1332+2634 & Sy (Sy) & 13:32:45.61 &  +26:34:49.2 & 14102 & 2015-01-10 & 438.5     & 77.9 & 30.60     &     3.6 &     2.4 &     1.4 \\
 25 &  J1347+4654 &  A (Sy) & 13:47:57.71 &  +46:54:35.0 &  8295 & 2015-04-23 & 594.4     & 69.0 & 58.09     &     5.5 &     4.1 &     1.3 \\
 27 &  J1402+0918 & Sy (Sy) & 14:02:28.72 &  +09:18:56.4 &  5731 & 2015-03-28 & 600.0$^d$ & 58.7 & 40.80$^d$ & ---$^d$ & 4.3$^d$ & ---$^d$ \\
 31 &  J1435+1007 & Sy (Sy) & 14:35:23.42 &  +10:07:04.6 &  9364 & 2015-01-23 & 584.6     & 60.9 & 41.94     &     4.1 &     2.5 &     1.4 \\
 36 &  J0021+0033 &  Sy (C) & 00:21:45.82 &  +00:33:27.3 &  5396 & 2006-02-05 & 585.6     & 51.2 & 45.43     &     4.6 &     2.8 &     1.4 \\
 50 &  J0907+3528 &  Sy (C) & 09:07:37.06 &  +35:28:28.5 &  8283 & 2014-11-05 & 438.6     & 86.1 & 80.36     &     9.3 &     6.0 &     1.4 \\
 53 &  J0917+1910 &   C (C) & 09:17:20.87 &  +19:10:19.3 &  8559 & 2014-11-14 & 584.6     & 65.9 & 46.40     &     4.7 &     2.9 &     1.4 \\
 62 &  J0956+0214 &   L (C) & 09:56:09.64 &  +02:14:48.4 &  9593 & 2006-03-10 & 579.8     & 53.1 & 30.88     &     3.0 &     1.8 &     1.4 \\
 67 &  J1021+6352 &   C (C) & 10:21:49.13 &  +63:52:06.8 &  6325 & 2014-12-16 & 585.2     & 50.1 & 58.89     &     5.7 &     3.4 &     1.4 \\
 89 &  J1159+5118 &   C (C) & 11:59:22.33 &  +51:18:09.0 &  8820 & 2015-04-22 & 596.7     & 39.3 & 66.08     &     5.8 &     3.7 &     1.3 \\
 92 &  J1220+0534 &  Sy (C) & 12:20:24.29 &  +05:34:21.9 &   937 & 2014-11-15 & 584.7     & 55.6 & 46.99     &     4.8 &     2.9 &     1.3 \\
 94 &  J1225+0519 &   C (C) & 12:25:05.60 &  +05:19:44.6 &  2044 & 2014-11-15 & 594.7     & 53.8 & 46.75     &     4.6 &     2.8 &     1.3 \\
102 & J1338--0023 &   C (C) & 13:38:15.40 & --00:23:54.7 &  6613 & 2015-01-11 & 584.6     & 51.1 & 40.03     &     3.9 &     2.4 &     1.4 \\
\hline
\end{tabular}
\tablenotes{
\item \textit{c} --- J1198+6123 was observed for 10 minutes of integration on 2015-03-12 under GBT session AGBT15A\_209\_11. Our re-reduction of the archival data for this galaxy showed a spectrum with a wavy baseline that was not well-fit by a 5th-degree polynomial. The MCP record of this galaxy lists values for a scan of $t_{\rm{Int}}=3$ min., though this appears to be rounded to the nearest minute from an actual integration time of 150 s, or the duration of a single GBT scan on this galaxy. We therefore take values for elevation, $T_{\rm{sys}}$, and $\Delta v$ from our re-reduction, and the integration time and $\rm{RMS}=6.3$ mJy from the MCP. \\
\item \textit{d} --- The archival data for J1402+0908, under project code AGBT15A\_209\_12, could not be accessed, so we show here the values maintained by the MCP at \url{https://safe.nrao.edu/wiki/bin/view/Main/MegamaserProjectSurvey}.
}
\end{table*}

\section{Masers in Low-mass NSA Galaxies} \label{sec:low-mass-masers}

\begin{figure*} 
    \includegraphics[width=0.98\linewidth]{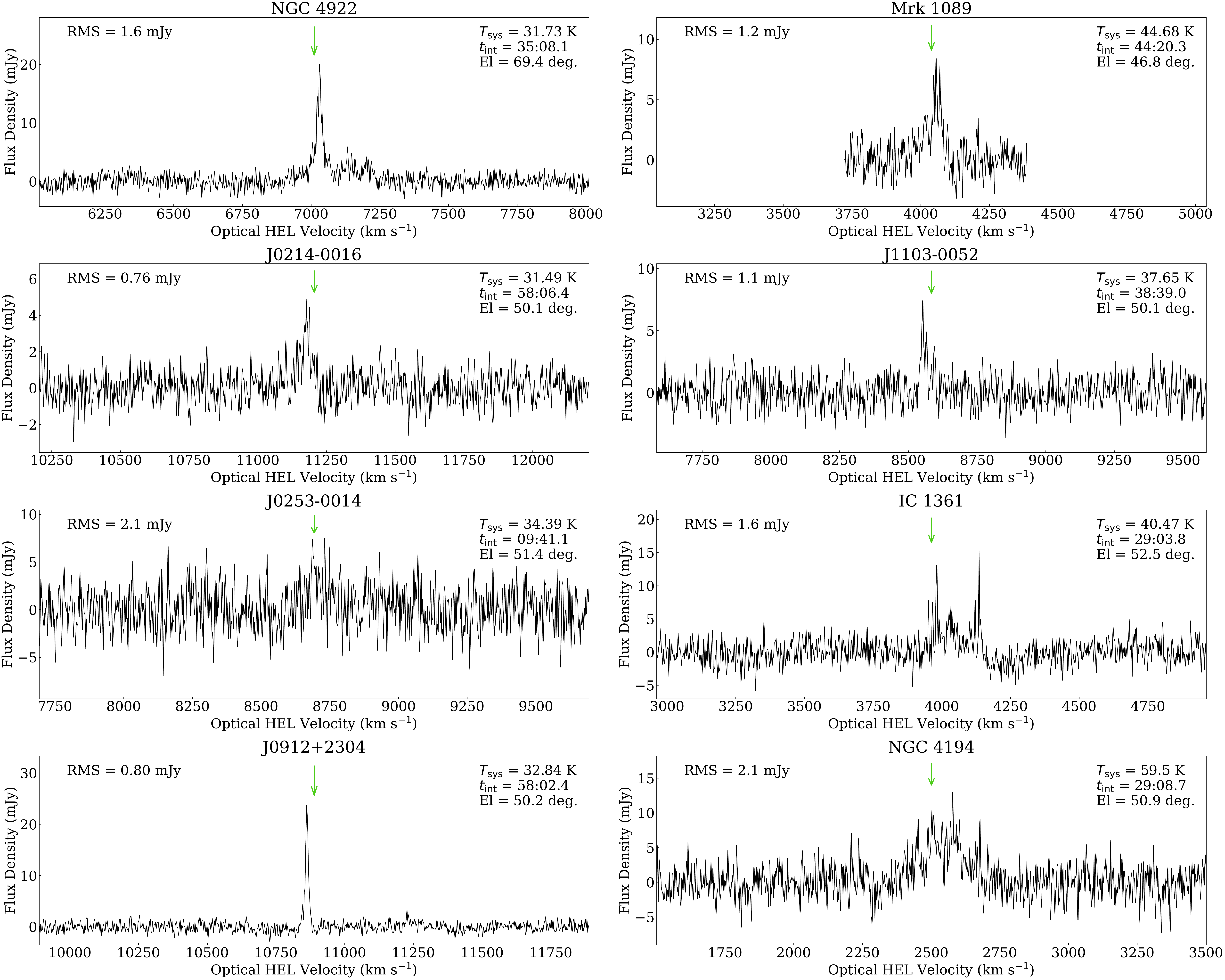}
    \includegraphics[width=0.70\linewidth]{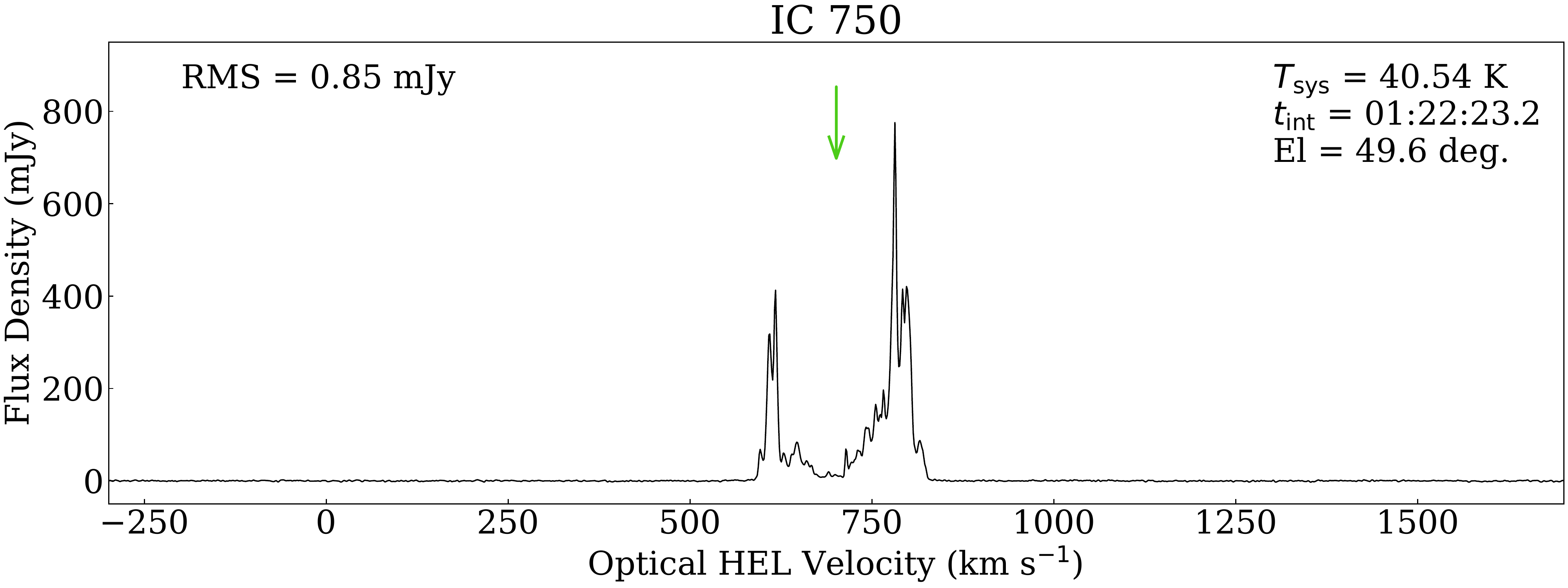}
    \caption{The GBT spectra of the eight known masers in low-mass galaxies with reliable distances in the NSA, and IC 750, currently the only confirmed disc maser in a low-mass galaxy \citep{Zaw20}. Data were reduced in an identical manner to other archival GBT data, as described in Section \ref{sec:data}. The average system temperatures, integration times, and elevations at the start of observation for the displayed spectra are shown int he top right corner of each panel. We show the line-free RMS noise, after both boxcar smoothing to $\sim$1 km s$^{-1}$ and Hanning smoothing, in the top left corner of each panel. In all panels, the systemic recessional velocity of the galaxy is marked by the green arrow. Galaxies are plotted in ascending order in $M_*$ (see Table \ref{table:lowmass-masers}).} \label{fig:lowmass-masers}
\end{figure*}

As stated in Section~\ref{subsubsec:mass}, water maser emission associated with AGN activity has been detected by previous surveys in eight low-mass galaxies with reliable distance measurements. One galaxy, NGC 4922, is a dwarf galaxy with a total stellar mass of $M_*=10^{9.31}~M_\odot$, and the remaining seven have total stellar masses between $10^{9.5}~M_\odot$ and $10^{10.0}~M_\odot$. We list the optical properties of these galaxies in Table \ref{table:lowmass-masers}. We re-reduced the archival GBT data for these maser systems using the same methods discussed in Section \ref{sec:data}. If there are multiple observations for a galaxy, we take the only the one with the lowest RMS noise, usually the longest observation. The spectra are shown in Figure \ref{fig:lowmass-masers}, and the dates of the corresponding observations are listed in Table \ref{table:lowmass-masers}. None of the spectra show the characteristic emission of a disc maser system, namely three emission complexes, one at the systemic velocity of the galaxy, one blueshifted by the rotational velocity from the systemic velocity, and one redshifted by the rotational velocity\footnote{In some systems, disc masers only have two out of the three complexes.}.

In Table \ref{table:lowmass-masers} and Figure \ref{fig:lowmass-masers}, we have also included IC 750. It is currently the only disc maser in a low-mass galaxy that has been confirmed with VLBI observations \citep{Zaw20}, and motivated our search for maser emission in active dwarf galaxies. This galaxy was not in the samples used in Section \ref{subsec:lumandmass} because it does not have a reliable distance in the NSA. The galaxy and AGN properties listed are taken from \citet{Zaw20} who used a secure distance measurement. The total stellar mass measured from {\it Spitzer} photometry, $M_* = 10^{10.1\pm 0.2}~M_\odot$ \citep{Zaw20}, is in good agreement with the NSA value of $M_* = 10^{9.94}~M_\odot$.

\begin{table*}
\caption{Water Masers Associated with AGN in Low-Mass Galaxies} \label{table:lowmass-masers}
\begin{tabular}{ccccccccccc}
    \hline \hline
    (1) & (2) & (3) & (4) & (5) & (6) & (7) & (8) & (9) & (10) \\
    Name & RA & Dec & $v_{\rm{sys}}$ & GBT Obs. Date & BPT & $\rm{H} \alpha / \rm{H} \beta$ & $L_{\rm{[O~III],obs}}$ & $L_{\rm{[O~III],corr}}$ & $\log~M_*$ \\
     & & & km s$^{-1}$ & & & & erg s$^{-1}$ & erg s$^{-1}$ & $\log~M_{\sun}$ \\
    \hline
    NGC 4922 & 13:06:25.26 & +29:18:49.5 & 7011 & 2004-01-29 & Sy & 7.2 & $4.1\times10^{40}$ & $5.5\times10^{41}$ & 9.31 \\
    Mrk 1089$^a$ & 05:01:37.76 & --04:15:28.4 & 4019 & 2010-02-01 & -- & -- & -- & -- & 9.59 \\
    J0214--0016 & 02:14:05.91 & --00:16:37.1 & 11205 & 2008-02-28 & Sy & 4.4 & $4.2\times10^{40}$ & $1.3\times10^{41}$ & 9.60 \\
    J1103--0052 & 11:03:38.36 & --00:52:08.6 & 8584 & 2011-01-07 & Sy & 6.1 & $5.3\times10^{39}$ & $4.3\times10^{40}$ & 9.78 \\
    J0253--0014 & 02:53:29.61 & --00:14:05.6 & 8693 & 2006-03-18 & Sy & 5.5 & $4.9\times10^{40}$ & $3.0\times10^{41}$ & 9.81 \\
    IC 1361$^b$ & 21:11:29.10 & +05:03:16.0 & 3962 & 2009-12-01 & -- & -- & -- & -- & 9.89 \\
    J0912+2304 & 09:12:46.37 & +23:04:27.4 & 10889 & 2008-01-31 & Sy & 4.4 & $4.4\times10^{40}$ & $1.4\times10^{41}$ & 9.98 \\
    NGC 4194$^c$ & 12:14:09.49 & +54:31:37.0 & 2501 & 2014-01-15 & -- & -- & $1.5\times10^{40}$ & -- & 9.98 \\
    \hline
    IC 750$^d$ & 11:58:52.20 & +42:43:21.1 & 701 & 2011-03-01 & Sy & 10.3 & $2.5\times10^{38}$ & $7.4\times10^{38}$ & 10.15 \\
    \hline
\end{tabular}
\begin{tablenotes}
    \item The galaxies in the table are listed in order of increasing total stellar mass.
    \item (1) --- Galaxy name as listed by MCP.
    \item (2) --- Right ascension (J2000).
    \item (3) --- Declination (J2000).
    \item (4) --- Heliocentric systemic velocity.
    \item (5) --- Observation date (year-month-day) for the corresponding spectrum shown in Figure \ref{fig:lowmass-masers}.
    \item (6) --- BPT classification, using both [N II]/H$\alpha$ and [S II]/H$\alpha$ diagnostics as described in Section \ref{subsec:BPT}. Possible values are: Sy--Seyfert, L--LINER, C--Composite, A--Ambiguous, SF--Star-forming.     
    \item (7) --- Balmer decrement.
    \item (8) --- Observed [O III] luminosity.
    \item (9) --- [O III] luminosity corrected for extinction using the Balmer decrement and the correction formula from \citet{Bassani99}.
    \item (10) --- Total stellar mass of the galaxy, taken from the NSA.
    \item $a$ --- There is no spectroscopic information for Mrk 1089 (also called NGC 1741) listed in the NSA.
    \item $b$ --- There is no spectroscopic information for IC 1361 listed in the NSA.
    \item $c$ --- NGC 4194 does not have good values of H$\alpha$, H$\beta$, or [N II] flux in the NSA.
    \item $d$ --- The distance listed in the NSA for IC 750 is unreliable, specifically $\rm{ZDIST/ZDIST\_ERR} < 10$. Therefore the Balmer decrement, [O III] luminosities, and stellar mass for IC 750 are taken from \citet{Zaw20}, which uses a secure distance measurement.
\end{tablenotes}
\end{table*}

%%%%%%%%%%%%%%%%%%%%%%%%%%%%%%%%%%%%%%%%%%%%%%%%%%

% Don't change these lines
\bsp	% typesetting comment
\label{lastpage}
\end{document}